\documentclass[reprint,aps,prl,twocolumn,balancelastpage,a4paper,floatfix,superscriptaddress]{revtex4-2}

\usepackage[margin=1in]{geometry}
\usepackage[T1]{fontenc}
\usepackage{graphicx}
\usepackage{float}
\usepackage{amsmath,amssymb,bm}
\usepackage{xcolor}
\usepackage{booktabs}
\usepackage{enumitem}
\usepackage{dcolumn}
\usepackage{hyperref}
\usepackage{hycolor}
\usepackage{setspace}
\hypersetup{colorlinks=true,citecolor=[rgb]{0,0.15,0.60}, 
urlcolor=[rgb]{0,0.15,0.60}, linkcolor=[rgb]{0,0.15,0.60}, final = true}
\usepackage[rightcaption]{sidecap}
\usepackage{physics}
\usepackage{mathtools}
\usepackage{relsize}
\usepackage{times}
\usepackage{siunitx}
\usepackage[scaled=0.86]{helvet}
\setcitestyle{round,super}

  \newcommand{\ie}{\textit{i.e.}}
  \newcommand{\eg}{\textit{e.g.}}
  \def\fref#1{Fig.~\ref{#1}}
  \def\sb#1{\textbf{\textsf{#1}}}
  \def\nsb#1{\noindent\textbf{\textsf{#1~}}}

  \definecolor{YKB}{rgb}{0.00,0.18,0.65}

\begin{document}

\title{\sb{\larger[1] Convergent evolution in a large cross-cultural database of musical scales}}

  \author{John M. McBride}
    \affiliation{Center for Soft and Living Matter, Institute for Basic Science, Ulsan 44919, South Korea}
    \email{jmmcbride@protonmail.com}
  \author{Sam Passmore} 
    \affiliation{Faculty of Environment and Information Studies, Keio University, Fujisawa, Japan}
    \affiliation{Evolution of Cultural Diversity Initiative, College of Asia and the Pacific, Australian National University}
  \author{Tsvi Tlusty}
    \affiliation{Center for Soft and Living Matter, Institute for Basic Science, Ulsan 44919, South Korea}
    \affiliation{Departments of Physics and Chemistry, Ulsan National Institute of Science and Technology, Ulsan 44919, South Korea}
    \email{tsvitlusty@gmail.com}



\begin{abstract}
\textsf{
{\setstretch{1.2}
{\larger[0.5] Scales, sets of discrete pitches that form the basis of melodies, are thought to be one of the most universal hallmarks of music. But we know relatively little about cross-cultural diversity of scales or how they evolved. To remedy this, we assemble a cross-cultural database (Database of Musical Scales: \href{https://github.com/jomimc/DaMuSc}{DaMuSc}) of scale data, collected over the past century by various ethnomusicologists. Statistical analyses of the data highlight that certain intervals (\eg, the octave, fifth, second) are used frequently across cultures. Despite some diversity among scales, it is the \textit{similarities} across societies which are most striking: step intervals are restricted to \SIrange{100}{400}{cents}; most scales are found close to equidistant 5- and 7-note scales. We discuss potential mechanisms of variation and selection in the evolution of scales, and how the assembled data may be used to examine the root causes of convergent evolution.}}}
\end{abstract}

\maketitle

\section*{\sb{Introduction}}
  Music, like language, can be described as a generative grammar consisting
  of basic building blocks, and rules on how to combine them.~\cite{ler96,pat10}
  In melodies, these basic units are usually specified by two quantities: frequency and duration. We generally refer to this basic pitch unit as a note, and a set of notes as a scale. Thus, as far as pitch is concerned, a scale is to a melody what an alphabet is to writing. 
  Despite their centrality to music and apparent ubiquity, we know surprisingly little about scales. Most studies focus on scales from a limited number of musical traditions,~\cite{bur99,car99} and the main statistical findings are that scales are non-equidistant, and have 7 or fewer notes.~\cite{khewm77,savpn15} There are many anecdotal reports that certain notes are commonly used,~\cite{kolet67,har76,bur99,car99,tre00,bropm13} but this has only been quantitatively examined in one small cross-cultural sample.~\cite{jiepf19}
  Thus, we lack concrete understanding of the fundamental questions of why humans use scales, how diverse they are, or how they came to be that way. We suspect that this is simply due to a lack of suitable resources.
  Here, we address this problem by presenting and analyzing a data set of musical scales from many societies, extant and extinct, built upon a century of ethnomusicological enterprise.

\begin{figure*}
\centering
\includegraphics[width=0.95\textwidth]{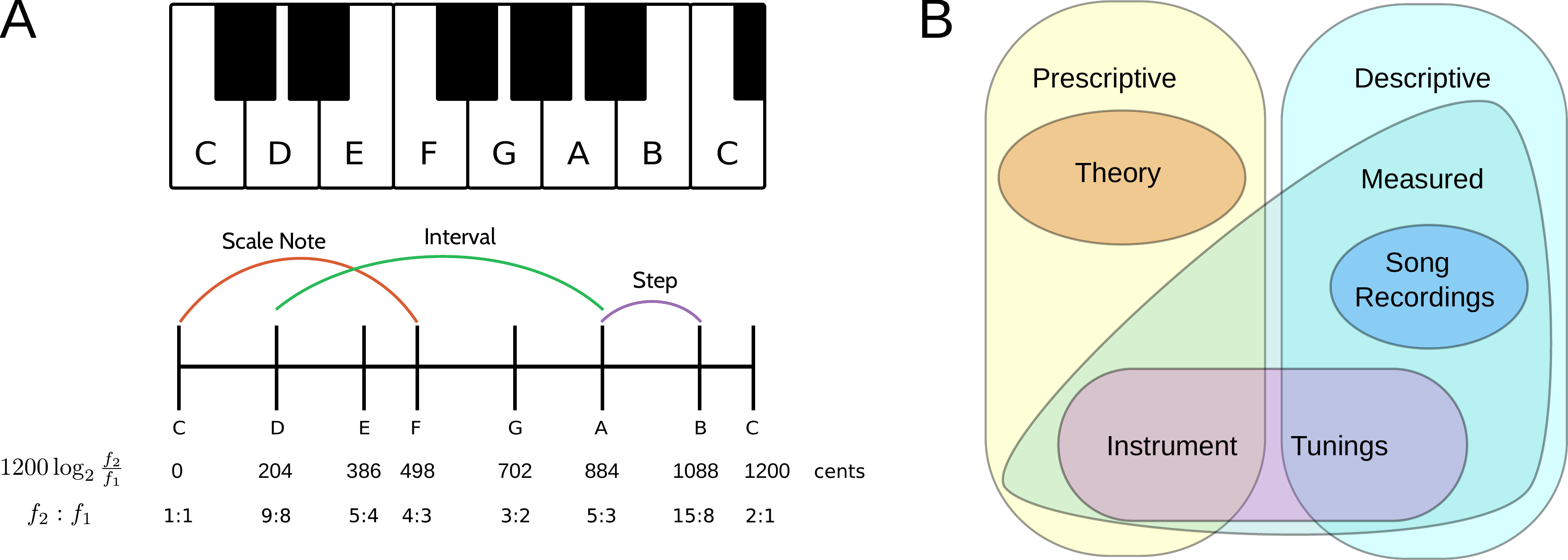}
\caption{\label{fig:intro} 
  \sb{Basic notions.}
  A: Illustration of relevant terms: Notes in a scale can be represented symbolically (\eg, letters for notes), or quantitatively. As examples, we show: the Western major scale in the key of C on a piano (top);   and the corresponding intervals in cents and as frequency ratios (bottom; in 5-limit just intonation tuning).
  We show three types of intervals: a scale \textit{note}, a \textit{step}, and an \textit{interval} that is neither a note or a step. 
  B: Venn diagram indicating how scale data can be classified.
}
\end{figure*}

  Let us begin by clarifying a few key terms and ideas. We define a \textit{scale} simply as a sequence of unique notes (Figure \ref{fig:intro}A). \textit{Notes} are pitch categories described by a single pitch, but as pitch is continuous, notes are more realistically described as regions of semi-stable pitch centered around a representative (\eg, mean, median) frequency.~\cite{pfojc17,sch20} However, humans process relative frequency much better than absolute frequency,~\cite{levtc05} so scales are practically specified by the \textit{intervals} (frequency ratios) between notes. Thus, a scale is usually defined by the ratios made between all notes and the first note, called the \textit{tonic}; we hereafter refer to these as \textit{scale notes}. Intervals are determined by their frequency ratios, but since we perceive pitch logarithmically,~\cite{attaj71,jaccb19} they are naturally
  measured in units of \textit{cents} --  obtained by a base-2 logarithm of the ratio of frequencies $f_1$ and $f_2$, $\textrm{cents} = 1200 \times \log_{2}f_1/f_2$.
  In what follows, \textit{steps} will describe the intervals between adjacent notes,
  and \textit{intervals} can refer to the interval between any pair of notes.

  We explicitly define an important sub-class of scales, \textit{octave scales}, which span an octave (frequency ratio 2:1; \SI{1200}{cents}) and have circular (or helical) structure such that notes repeat at every octave;~\cite{shepr82}
  \eg, \fref{fig:intro}A shows that the first and last notes (an octave apart) of such a scale have the same name -- a property is called \textit{octave equivalence}). It is often reported that octave equivalence is one of the most universal features of music.~\cite{kolet67,har76,bur99,car99,tre00,bropm13}   However, this has been disputed,~\cite{udoma97,net00,net10,jaccb19,sch20} and statistical study of octave prevalence is still lacking. Octaves are indeed salient in the harmonic spectra.~\cite{llojr42,parmp18} Yet, experiments have indicated that octave equivalence is only a weak perceptual phenomenonon,~\cite{dowpp77,kruco00,bonje13,demhr21} that depends heavily on musical training,~\cite{allps67,kalpp82,hoeap12} and culture.~\cite{jaccb19} Nonetheless, several measurable phenomena related to harmonic tones may lead to preferential use of the octave: tonal fusion,~\cite{stu90,mcpnc20,demhr21} hearing in noise,~\cite{mcpap22}, and memorability of complex tones.~\cite{mcppn20} For example, increased tonal fusion between octave intervals (likewise for fourths, fifths, which also undergo tonal fusion) may give rise to
  greater perception of synchrony when singing in parallel.~\cite{savbb21}

  To address longstanding questions on scale diversity and evolution, we first document
  the creation of a database of scales. We then analyse the empirical scales to see which
  intervals are significantly frequent or infrequent, providing robust statistical evidence
  that the octave is prevalent among many societies. We examine the cross-cultural diversity
  of octave scales, finding that within-society variation can in some cases be as great
  as the total variation in scales -- scales are surprisingly not that diverse.
  In particular, across all geographical regions step intervals are limited to \SIrange{100}{400}{cents}, and 
  scales are clustered around 5- and 7-note equidistant scales (scales where step sizes
  are similar in size). As a result of these restrictions, humans tend to use only $\sim$\SI{1}{\%}
  of possible octave scales, which strongly suggests some degree of convergent evolution due to
  shared biases (in addition to convergence via cultural diffusion).
  Finally, we discuss the potential mechanisms of change and selection of scales,
  discuss the challenges in understanding scales evolution,
  and propose credible future directions for studying this evolution.

\section*{\sb{Scales Database}}

\nsb{Database curation.}
  A total of 60 books, journals, and other ethnomusicological sources were found to
  have relevant data on scales (SI Table 1).~\cite{hew13,rec18,ell85,wacna50,kubam64,bra67,kun67,rousf69,traam70,traam71,kni76,sur72,mor76,haee79,kubam80,vanam80,zeme81,anija82,hoam82,kubam84,kubyt85,got85,yuapp87,keemp91,traam91,carja93,sche01,attpi04,zhaa04,set05,mcnja08,str09,kus10,garan15,wis15,moric18,bad19,kim76,nammp98,weiae13,ambwm05,ambms06,cooyi71,garam75,jonet66,jonam60,kubam62,kubam63,kubwm92,thram78,min90,sunst69,copef18,mil85,sch20,mzhmu20,mokna37,tan11,men12,panjn13} 
  We note one previous attempt to create a database of scales from the ethnomusicological  literature,~\cite{honjn11} however there are no details on its construction, and it does not link scales directly to original sources.
  Our database improves on these issues through a stringent methodology (described below), sources for all scales and is presented as a digitized, open-access, resource which will continue to grow as new data is made available.

  We can define scales (\fref{fig:intro}B) either prescriptively ("these are the notes you can use in a melody") or descriptively ("these are the notes that were used in the melody"). 
  \textit{Theory} scales consist of intervals with idealized, exact frequency ratios -- they lack the random fluctuations that exist in the real world. These are mainly found in a limited set of cultures that exist along the old Silk Road route, and they are not necessarily played according to these theoretical ideals.~\cite{sig77,panjn13}
  Absence of theory scales does not mean that other societies lack musical theory, which can be found either explicitly through language, or implicitly through an understanding of what is `correct' and `incorrect' in a particular style.~\cite{zemet79,ricyi80,felyt81,baiyt88,pat10,chu15} Theory scales are by definition prescriptive scales, although descriptive scales can be found which closely match these.
  \textit{Measured} scales are obtained where measurements have been made of the notes on an instrument, or a recording of a song has been analysed with computational tools to extract a scale. Instrument tunings are by default prescriptive, but can be descriptive if all notes are used in a melody. There is some error in these measurements -- taken with tools that include tuning forks, the Stroboconn, and modern computational approaches -- but it is always within \SI{10}{cents}.
  Measured scales taken from song recordings are exclusively descriptive, and they make up the smallest part of the database. This is because it is still quite a challenge to reliably infer scales from a recording of a performance using algorithms,~\cite{ambms06} and thus it requires extensive manual labor. Still, we believe that the future of studies on musical scales lies in tackling these challenges, due to the prospect of extracting descriptive scales from archives of ethnographic recordings.~\cite{corsp10,rosti20,woopo22,sixjn13} However, it is clear that the appropriate computational tools to perform such large-scale analyses are still lacking.~\cite{corjn13,benja15,ozais21}

  \begin{figure*}
  \centering
  \includegraphics[width=0.95\linewidth]{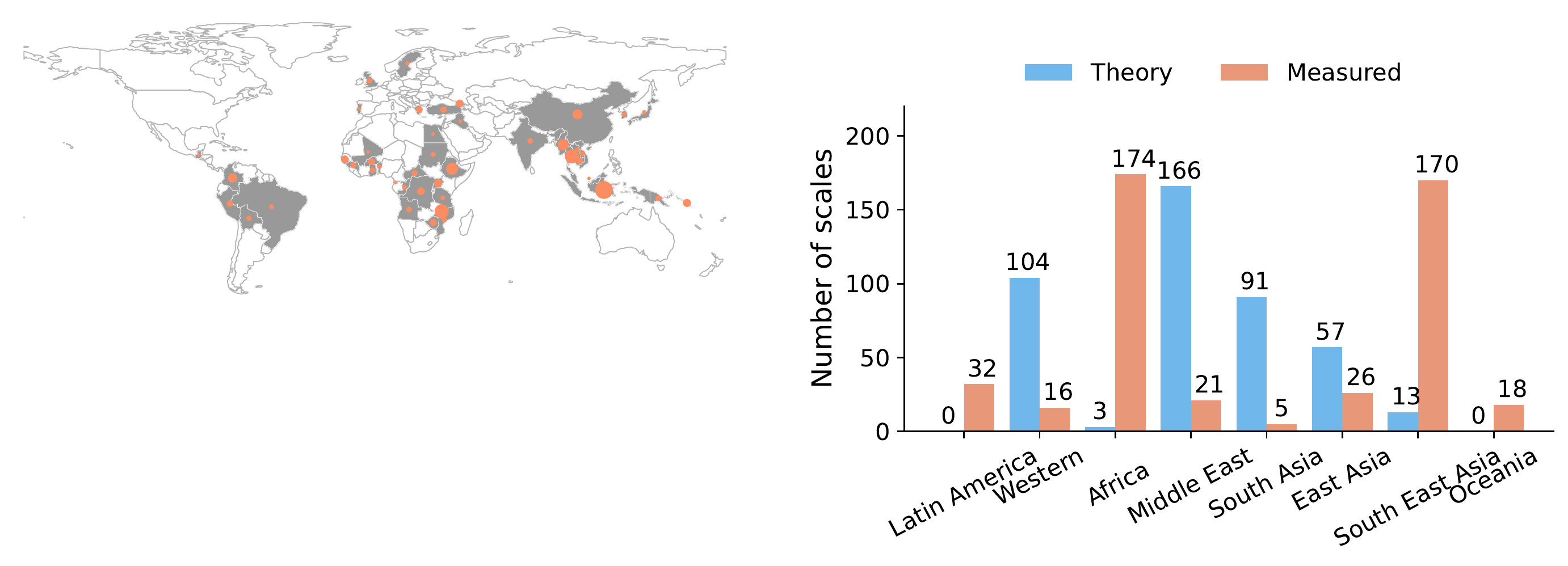}
  \caption{ \label{fig:world}
  Breakdown of scales in the database according to: geographical area;
  theoretical scales or measured scales. The map shows
  the geographic origin of the measured scales; theory scales are not included
  since they are not always associated a single country; marker size indicates
  sample size. Made with Natural Earth.~\cite{geo20}
  }
  \end{figure*}

  Another statistical regularity in musical pitch is \emph{tonality} -- \ie, the distribution of notes, how they are played (duration, ornamentation), and their transition probabilities.~\cite{savcb22,verpm21} Of the various aspects of tonality, tonal hierarchies (probability distribution of notes)~\cite{kruje81} have been best documented across cultures.~\cite{casje84,kesmp84,nammp98,ambji09,harhs21}  Tonality (or modality) is in some cases an integral part of a scale that determines how the notes should be played,~\cite{peama16} to the extent that musicians can reliably tell apart two ragas from the duration of a single note.~\cite{ganja19}
  We emphasise that our definition of scale does not imply any specific tonal hierarchy, while acknowledging that similar definitions (\eg, maqam, raga) are often used with the expectation that they include tonality.~\cite{chu15} 
  In the database, we account for tonality in a limited way by noting the order of notes in the scale, starting from the first note (tonic). However, this information is  often unavailable, and we lack the finer details of note distribution or transition probabilities. Future cross-cultural analyses of recordings ought to document these important details.~\cite{peama16,ganti18}
  
  To enable a broad range of analyses we collected information pertaining to, where applicable, (i) the society (country, language or ethnic group, musical tradition), (ii) geography (country, geographic region) (\fref{fig:world}), (iii) instrument type, (iv) tonic note, and (v) whether the scale is measured using harmonic or melodic intervals.
  We additionally linked societies, where appropriate, to identifiers from other ethnographic and linguistic databases, such as D-PLACE and Glottolog.~\cite{kirpo16,ham19} D-PLACE contains information on social structure, economic, and environmental information, while Glottolog can be used to determine the linguistic distance between two groups - a common proxy for cultural similarity.~\cite{evapt21}   Some musical traditions that span multiple countries (such as Western classical music) are also taken as a unit of society. Identification of the society or geographic origins of a scale was required for inclusion of scale data. Other details were found at varying frequency; \eg, tonic was only identified in \num{121} out of \num{434} measured scales.\\

\nsb{Inferring octave scales.}
  The database exists in two forms: the raw data (434 theory scales, and 462 measured scales),
  and a set of octave scales that is generated procedurally from the raw data according to five choices.
  For a complete workflow from source to database, including examples, and the choices we made to
  generate octave scales, see SI section "Inferring octave scales, SI Fig. 1-3, and SI Table 1-2.

  In total, we infer \num{896} octave scales (\num{434} theory octave-scales, \num{384} octave-scales from instrument tunings,
  and \num{78} octave-scales from song recordings), from \num{73} societies. The theory scales span \num{6} regions, while the
  measured scales span \num{8} regions, and \num{46} countries (\fref{fig:world}).
  When inferring octave scales from measured scales, a principle assumption is that they add up to an octave.
  Since the validity of the assumption is not clear, we first study the statistics of the measured scales before studying octave scales.

\section*{\sb{Results}}

  \begin{figure*}  \centering
  \includegraphics[width=0.99\linewidth]{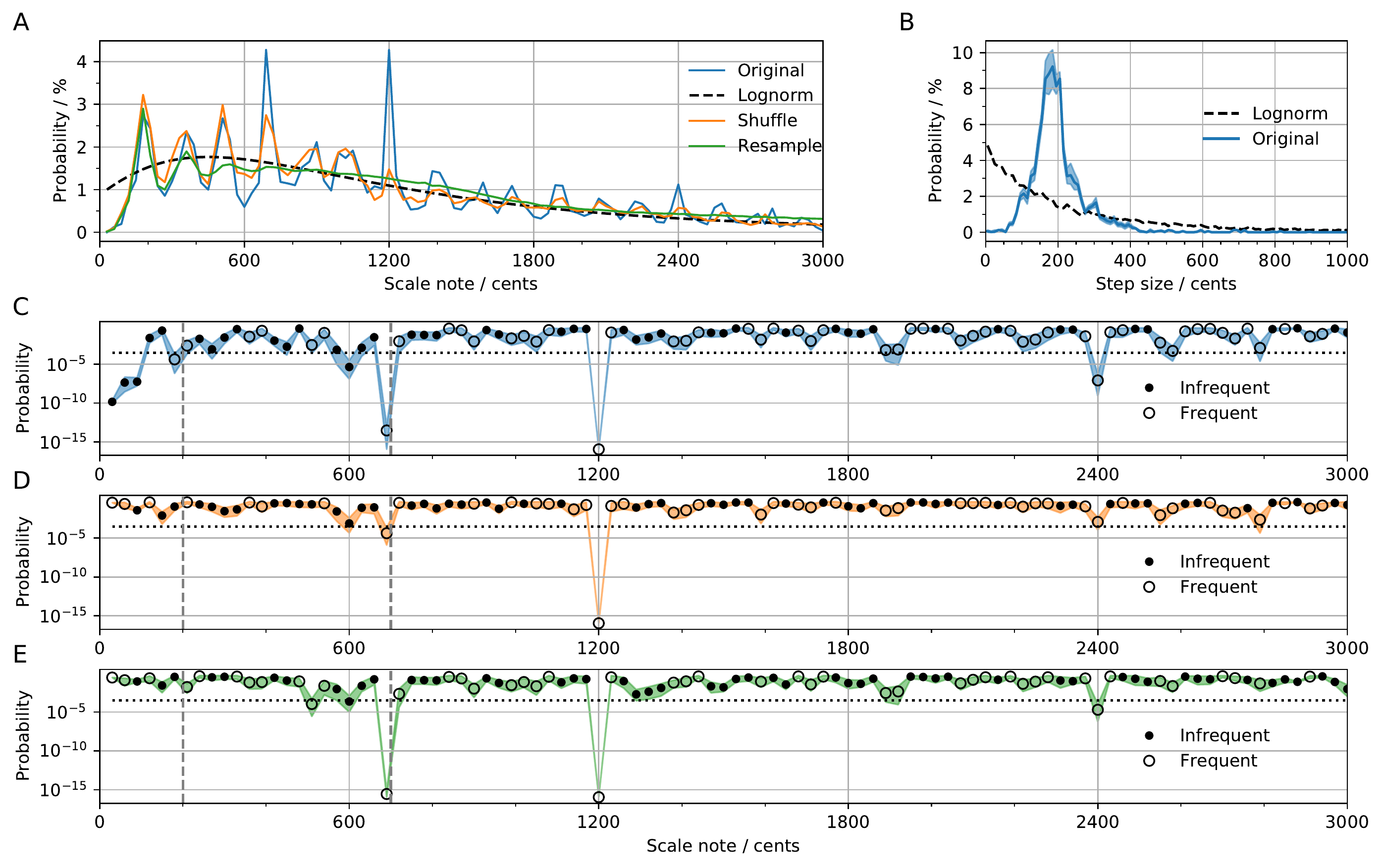}
  \caption{\label{fig:fig3}
  Between-scales distributions of step sizes and scale notes.
  A: Distribution of notes in measured scales (blue; histograms are shown as lines; bin size = 30 cents).
  Maximum likelihood lognormal distribution fitted to notes distribution (black).
  Distribution of notes obtained via alternative sampling:
  shuffling the step sizes within scales (orange); resampling step sizes
  from the full distribution of step sizes (green).
  The x-axis is truncated at \SI{3000}{cents} for clarity.
  B: Distribution of step sizes (blue). Distribution of step sizes
  in a set of scales with notes generated independently from a lognormal distribution.
  C-E: Probability that the counts observed in the original data were generated by
  the lognormal distribution (C), the shuffled scales distribution (D), or the resampled scales distribution (E).
  Empty circles indicate that the interval is found more than chance, while filled circles indicate the opposite. 
  Dotted line indicates a $p$ value of 0.05 after applying a Bonferroni correction.
  Dashed lines indicate the values of \SI{200} and \SI{700} cents.
  Shaded region shows bootstrapped $95\%$ confidence intervals.
  }
  \end{figure*}

 \noindent  One of our goals is to estimate whether certain notes and intervals appear more or less frequently than expected by chance. This is hard to quantify, since we lack a universally-correct method of calculating the probability of observing an interval. Instead, we propose three independent statistical models that reflect different ways of constructing scales:

  \nsb{Model: Lognorm.} This model assumes that scale notes $S$ are chosen independently
  from a lognormal distribution, $P(S) = \textrm{lnN}(\mu, \sigma^2)$.
  The rationale behind this choice is that small intervals should be uncommon
  ($P(S) \to 0$ as $S \to 0$) due to limits in pitch perception, and large intervals
  should be uncommon ($P(S) \to 0$ as $S \to \infty$) due to physical constraints (human anatomy; instrument size).
  We expect that this model is inappropriate for within-scale analyses, since notes within
  a scale are unlikely to be chosen independently of each other, but rather according to specific
  choices made by a musician. However, if the choices made by many independent musicians are sufficiently diverse,
  then the lognormal distribution may be an appropriate description of the between-scales note distribution.

  \nsb{Model: Shuffle.} This model assumes that specific step sizes are important, but their arrangement is not. Normally, both step size and order determine scale notes. But now consider a musician who cares only about the step sizes, not their order in which they are arranged. This can be approximated by sampling from the original
  data and reshuffling the order of step sizes.

  \nsb{Model: Resample.} This model assumes that step sizes are chosen independently from a distribution, and arranged randomly into scales. This is equivalent to a musician who is indifferent to both the step size, and the order. A reasonable choice of a distribution in this case is the posterior distribution of all step sizes.

  In any alternatively-sampled set of scales, the number of scales and the number of notes in these scales matches the original.
  In the following sections, we will use these three models to, first, compare intervals between scales, and second compare intervals within scales, searching for evidence of statistically significant counts of intervals.

  \begin{figure*}  \centering
  \includegraphics[width=0.99\linewidth]{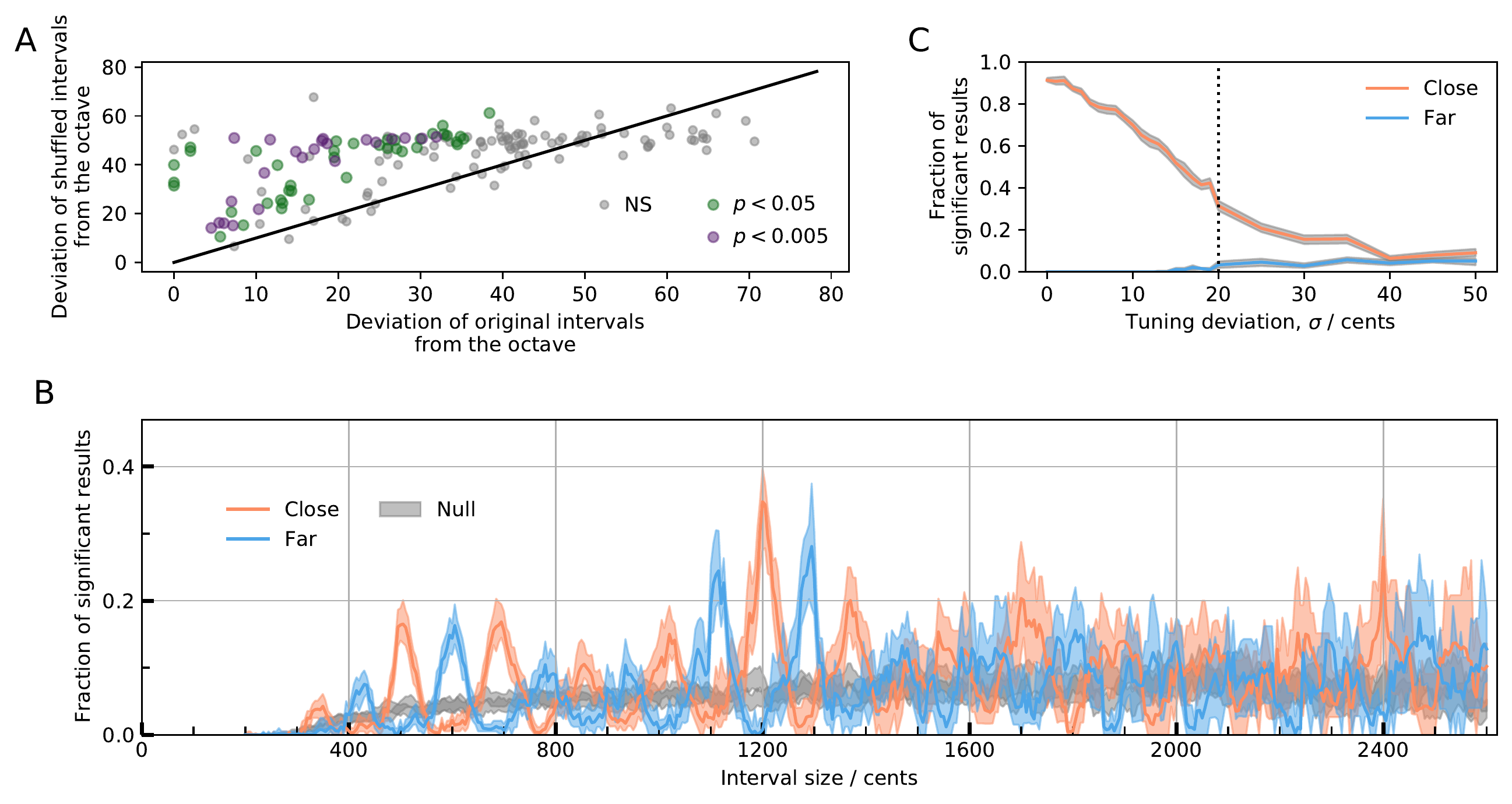}
  \caption{\label{fig:fig4}
  Within-scales tests of interval significance.
  A: Mean deviation of original intervals from the octave compared to shuffled intervals for \num{162} measured scales; colour indicates statistical significance; black line indicates $x=y$.
  B: Fraction of tested scales in which intervals are found to be significantly closer or further from
  a particular size than expected by chance. Null distribution (grey region) is obtained by repeating the analysis with resampled sets of scales.
  C: Fraction of significant results indicating that the intervals are closer or further to
  the octave than chance, against intonation error $\sigma$, in a test set which maximises
  the number of octave intervals. Dotted line indicates the point at which \SI{35}{\%}
  of results are significant.
  Shaded region shows bootstrapped $95\%$ confidence intervals.
  }
  \end{figure*}

\nsb{Statistically significant intervals between scales.}
  To search for statistically significant (due to an abundance or dearth of) intervals,
  we first plot the distribution of scale notes (\fref{fig:fig3}A, Original) and step sizes (\fref{fig:fig3}B, Original) for a sub-sample including \textit{only} measured scales,
  without imposing octave equivalence, and controlling
  for society (\num{214} scales sampled from a total of \num{434} scales, with no more than \num{5} per society, resampled \num{1000} times to achieve convergence; SI Fig. 4).
  We compare this empirical distribution to the maximum-likelihood lognormal distribution (\fref{fig:fig3}A, Lognorm),
  and the corresponding step size distribution (\fref{fig:fig3}B, Lognorm). As expected, we find that scales drawn from the lognormal distribution
  have step sizes that do not resemble real scales -- since notes within a scale are not independent.
  Instead, notes within scales are more evenly distributed, such that step sizes are peaked at ${\sim}\SI{200}{cents}$, and rarely smaller than \SI{100}{cents}
  (\fref{fig:fig3}B); this is found in every geographical region that we investigated (SI Fig. 5).
  However, scale notes from \textit{different} scales are independent of each other, resulting in a distribution that is approximately lognormal when many scales are sampled (\fref{fig:fig3}A).

  The utility of fitting a lognormal distribution to the data is that it can serve as a null-hypothesis baseline for estimating the probability that a scale note appears more or less than chance. To this end, we integrate the lognormal distribution over the range of each histogram bin $i$ in \fref{fig:fig3}A to get the probability of observing scale notes, $p_i$.
  We then calculate the binomial probability, $q_i = \binom{n}{k_i} p_i^{k_i} (1 - p_i)^{n-k_i}$, where $k_i$ is the number of observations in bin $i$, and $n$ is the total number of observations. We report either: the probability that $k_i$ or higher is observed if $k_i / n > p_i$,
  $\sum_{j=i}^{n} q_i$; or else, the probability that $k_i$ or less is observed, $\sum_{j=0}^{i} q_i$, if $k_i / n \leq p_i$.
  Low probability implies significant deviation from the null lognormal hypothesis.
  We see that only a few intervals deviate significantly from the lognormal distribution (\fref{fig:fig3}C):
  \SI{1200}, \SI{700}, \SI{200} and \SI{2400}{cents} are found more frequently, while \SI{600}{cents} is found less frequently than chance.
  This is strong evidence that the octave (and the fifth) are important intervals in many societies. 
  
  To corroborate these findings, we repeat the significance test with different assumptions on how scales are generated. We first repeatedly shuffle the step sizes in each scale to generate new scales, to examine whether the statistically significant intervals would arise if step sizes were ordered randomly (\fref{fig:fig3}D).
  Similarly, by resampling with new step sizes, we can test whether the significant intervals could plausibly have been produced by arranging randomly selected step sizes from the distribution in \fref{fig:fig3}B (\fref{fig:fig3}E).
  These analyses demonstrate that the peak at \SI{200}{cents} (\fref{fig:fig3}C) is due to it being the most common step size. The values of \num{600} and \SI{2400}{cents} are found to be less significant than in \fref{fig:fig3}C, but are still much more significant than most intervals.
  The fact that the peak at \SI{700}{cents}
  is significant in \fref{fig:fig3}D but not in \fref{fig:fig3}E is likely due to the presence of equidistant scales (SI Fig. 6), where the order of the intervals is mostly irrelevant. By this logic, one might expect the peak at \SI{1200}{cents} to disappear in \fref{fig:fig3}D, but it survives because a large fraction of equidistant scales do not extend beyond the octave (SI Fig. 6).
  Remarkably, the octave is extremely significant according to all three tests.

\nsb{Statistically significant intervals within scales.}
  The previous method was used to assess whether intervals are found more than expected by chance across a collection of scales. Since grouping together scales from different societies might be problematic, we require a test that can discriminate unusually frequent / rare intervals within individual scales. To this end, we compare the original scales with many alternative scales that are created by shuffling the original scales' step sizes.
  We use the full range of intervals that can be generated with an instrument, not just the scale notes. For an instrument with $N$ notes, this results in $N\times(N-1)/2$ intervals. By sampling many more intervals, this test is powerful enough to detect, in some cases, significant signals within individual scales.
  
  To test for statistical significance of an interval $I$, we find all intervals in a scale that fall within $w=\SI{100}{cents}$ of $I$, and calculate their distance from $I$. We repeat this process for \num{50} shuffled versions of the scale, collecting all intervals within $w$ of $I$ in a second group. We then use a Mann-Whitney U test to examine whether either set of intervals, original or shuffled, is significantly closer (\ie, $p<0.05$) to the octave than the other (repeating \num{100} times to get a converged average). We demonstrate this procedure for $I=\SI{1200}{cents}$ (\fref{fig:fig4}A): in most scales the original intervals are closer to an octave than the shuffled scales, although only a fraction of results are significant. 
  Reasons for non-significant results include small sample sizes and the tendency of equidistant scales to produce similar intervals when shuffled.

  We extend this analysis to all intervals over the range $200 \leq I \leq 2600$ cents, showing the fraction of significant results indicating an interval is found more or less frequently than chance (\fref{fig:fig4}B). Without a doubt, the most significant interval is the octave (\SI{35}{\percent} significantly close),  and the intervals that are most significantly avoided in scales are the regions flanking the octave. In further agreement with the between-scales analysis, the next most significant regions are those around \SI{500},
   \SI{600}{cents}, and \SI{700}{cents}.

  To put these results in perspective, we repeat the test on sets of scales generated by resampling step sizes from \fref{fig:fig3}B (\fref{fig:fig4}B, Null). The null distribution converges to \SI{5}{\%} as expected, which clearly demonstrates that the high fractions of significant results in the original scales are not artefacts due to testing multiple hypotheses.
  Additionally, we show that the results do not depend on our choice of $w$ (SI Fig. 7).
  The consistency of these test results reinforces the conclusion that these significant intervals are not chosen randomly.

\nsb{Effect of tuning variability on the search for universal intervals.}
  Observation of significant intervals likely reflects the intentions of the musician to tune to these intervals. However, non-significant results do not necessarily indicate lack of such intentions, as imprecision in tuning (and tuning measurements) may result in false negatives. To understand the impact of imprecision, we generate a test set of scales (by sampling step sizes from \fref{fig:fig3}B) and fix all scale notes that are $\SI{\geq 1200}{cents}$ to be exactly an octave higher than one of the notes that are $\SI{\leq 1200}{cents}$. We then add to the test sets normally-distributed noise, $\textrm{N}(\mu=0, \sigma^2)$. 
  Even without noise ($\sigma=0$), this test can find significant results only \SI{90}{\%} of the time (Figure \ref{fig:fig4}C),
  which demonstrates the difficulty in inferring intentionality simply due to low sample sizes.
  
  To estimate reasonable bounds on the noise $\sigma$, one may first consider that the measurement error varies from about  \SI{1}{cents} for computational methods (\eg, Stroboconn),~\cite{jonam70} to \SI{5}{cents} for tuning forks. Also, instruments vary in the stability of their notes: \eg, humans typically sing with a standard deviation of at least about \SIrange{10}{20}{cents}.~  \cite{hagst80,jerts05,ambji09,devpm11,seris11,pfojc17,damjv20,sch20} However, the main source of error is perception: humans make mistakes in discriminating intervals below $\SI{\sim 100}{cents}$ (although we lack replications with non-Western participants).~\cite{siepp77,mcdja10,zarja12,larbr19}
  Thus, we can expect a reasonable upper bound to the proportion of significant results
  that can be detected of about \SI{30} to \SI{70}{\%}. Thus, when we find the octave
  to occur significantly more than chance about $35\%$ of the time (\fref{fig:fig4}B), this
  is a rate that is within the range of the hypothetical maximum.
  
  Ultimately, it is hard to say how exactly scales are chosen.  Were some important intervals fixed first (\eg, the octave), and then the rest chosen to fill the gaps? Or were step sizes of a certain size (\eg, big and small size categories) chosen, and then arranged in some preferred order?~\cite{stu12}
  What we have conclusively shown is that independently of how scales are created, they show a significant, consistent bias towards including some notes, and avoiding others.\\

  \begin{figure}  \centering
  \includegraphics[width=0.99\linewidth]{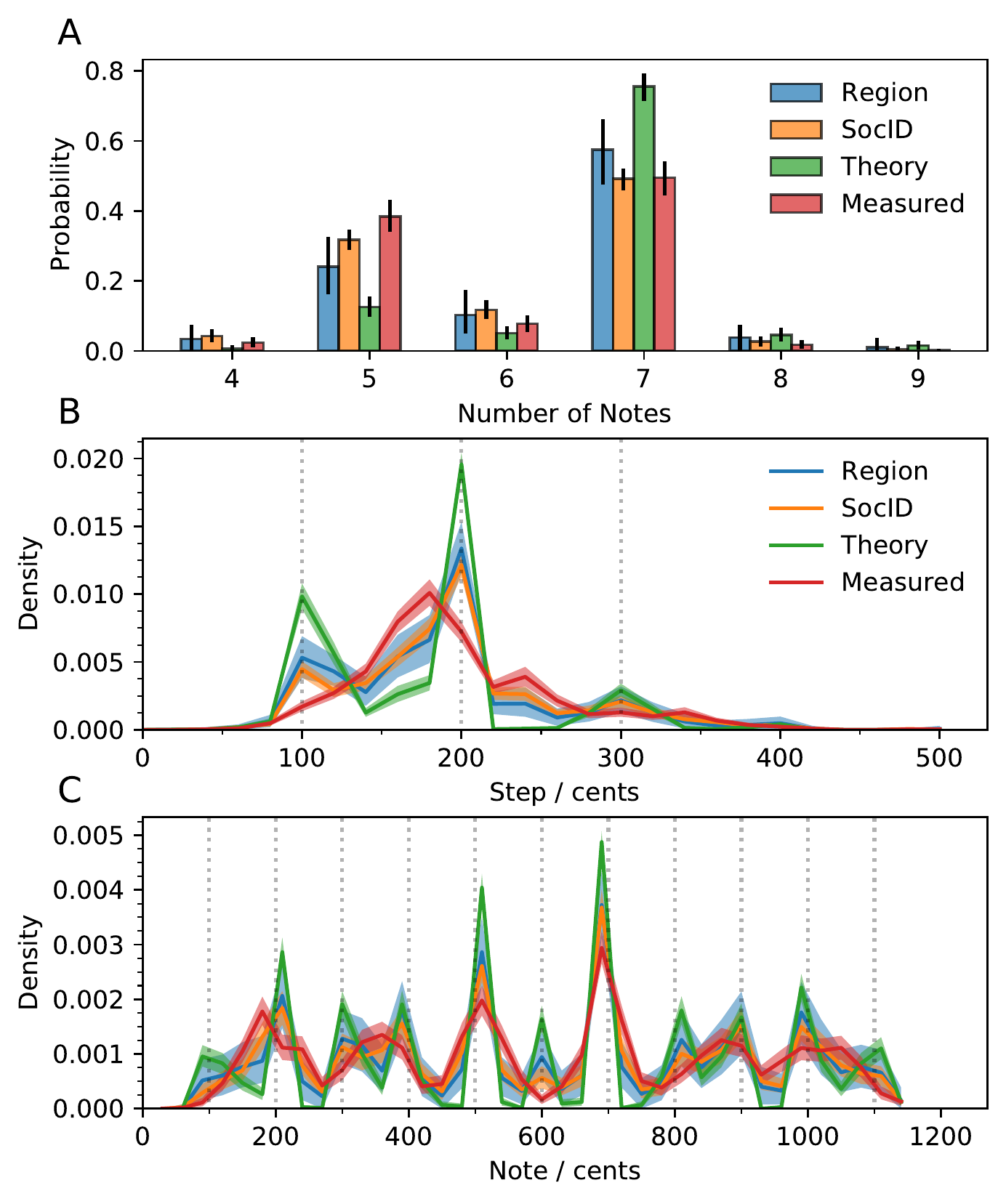}
  \caption{\label{fig:fig5}
  Statistics of octave scales.
  A: Distributions of the number of notes in a scale for different samples of the scale database: sample balanced by region, sample balanced by culture, theory scales, and measured scales.
  B-C: Histograms of step sizes (B) and notes in scales (C) for different samples of the scale database: samples balanced by region, samples balanced by society.
  Whiskers (A) and shading (B,C) indicate bootstrapped $95\%$ confidence intervals. Histogram bins are 20 (B) and 30 (C) cents. 
  }
  \end{figure}

  \begin{figure*}  \centering
  \includegraphics[width=0.99\linewidth]{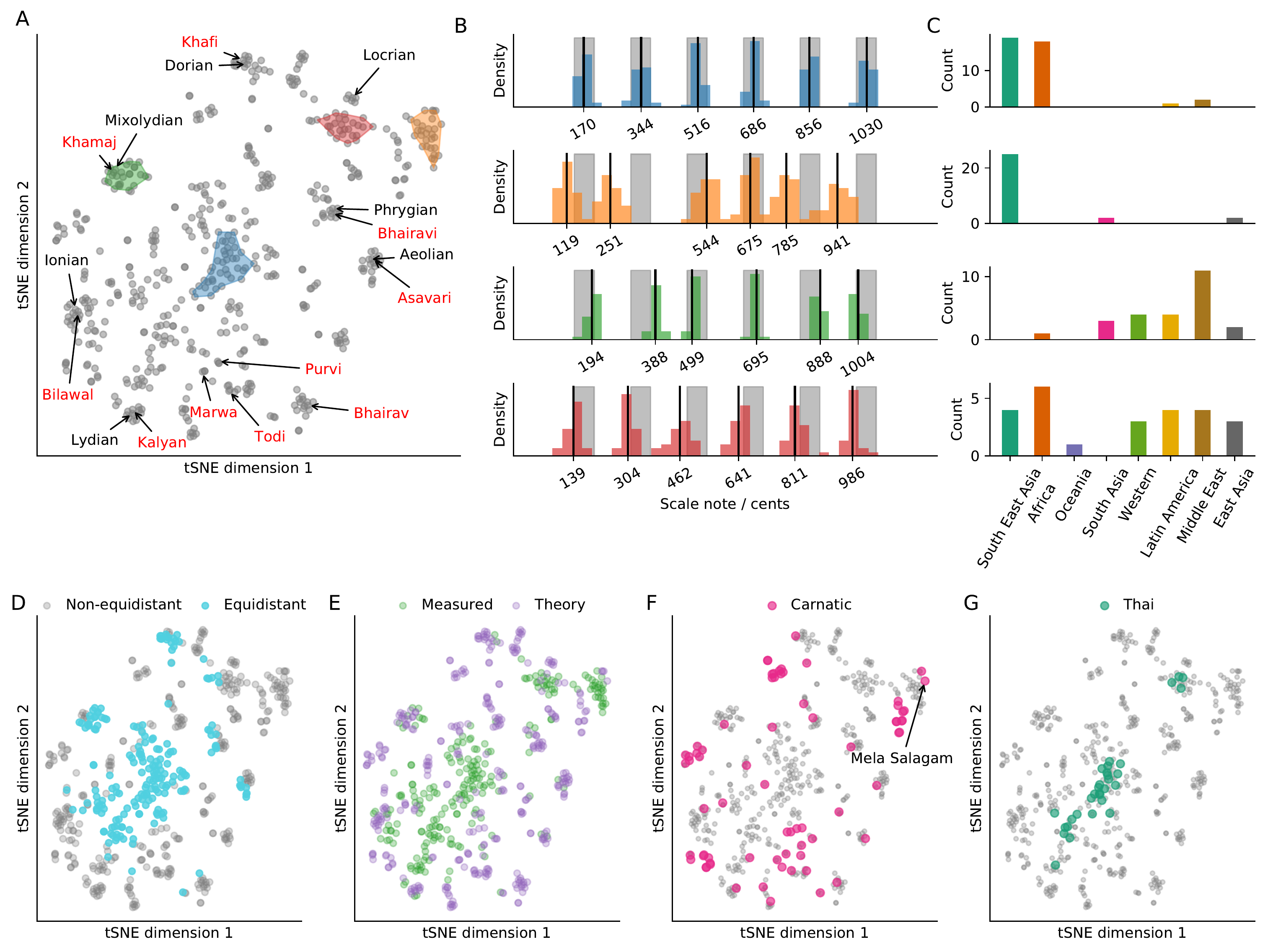}
  \caption{\label{fig:fig6}
  Cross-cultural diversity of scales.
  A: 2-dimensional embedding of \num{556} heptatonic scales, labelled with 7 Western diatonic modes (black),
  10 North Indian thaat (red), and the four largest clusters (shaded areas).
  B: Note distributions for each cluster. Black lines are shown for means of each note,
  and grey shading indicates equidistant scale notes $\pm$\SI{30}{cents}.
  C: Geographic distribution of each cluster.
  D-G: Embeddings are labelled with: equidistant scales (D; where notes are on average
  within \SI{30}{cents} of the equidistant values), theory vs measured scales (E),
  Carnatic scales (F), Thai scales (G).
  }
  \end{figure*}

\nsb{Qualitative evidence for preferential use of octaves.}
  A more direct route to understanding the intentions of musicians is through detailed ethnography. We therefore examined each source for qualitative evidence indicating preferential use (or absence) of the octave (SI Dataset 1). We found qualitative evidence in support of octave use in 26 out of 60 sources:
  use of the octave to tune instruments (8 sources); performing melodies in parallel octaves (11 sources);
  using the same name for notes an octave apart (10 sources), which is also strong evidence for octave equivalence.
 We also find quantitative evidence (statistically significant octaves for at least one scale; \fref{fig:fig4}A) in 26 sources;
  in total, 40 sources contain either quantitative or qualitative evidence. 
  For 6 of the remaining 20 sources, there is evidence from secondary sources for preferential use of the octave in those cultures;
  another 3 sources report on archaeological findings, which are difficult to investigate further.~\cite{haee79,yuapp87,zhaa04}
  
  Two sources (Georgian polyphonic singing) provided qualitative evidence that fifths are perceptually important, instead
  of octaves;~\cite{sch20,mzhmu20} however, we found statistically significant octaves in one of these sources, so the evidence here is mixed.~\cite{sch20}
  A third source (Colombian marimba) provided quantitative evidence that octaves were found significantly less than chance.~\cite{min90}
  Overall, we find some evidence (either primary or secondary) in support of octaves being significant
  in 46 out of 60 sources, and evidence to the contrary in 3 sources. These findings strongly support the view that octaves are widespread, but not absolutely
  universal.~\cite{net00,net10}\\

  \begin{figure*}  \centering
  \includegraphics[width=0.99\linewidth]{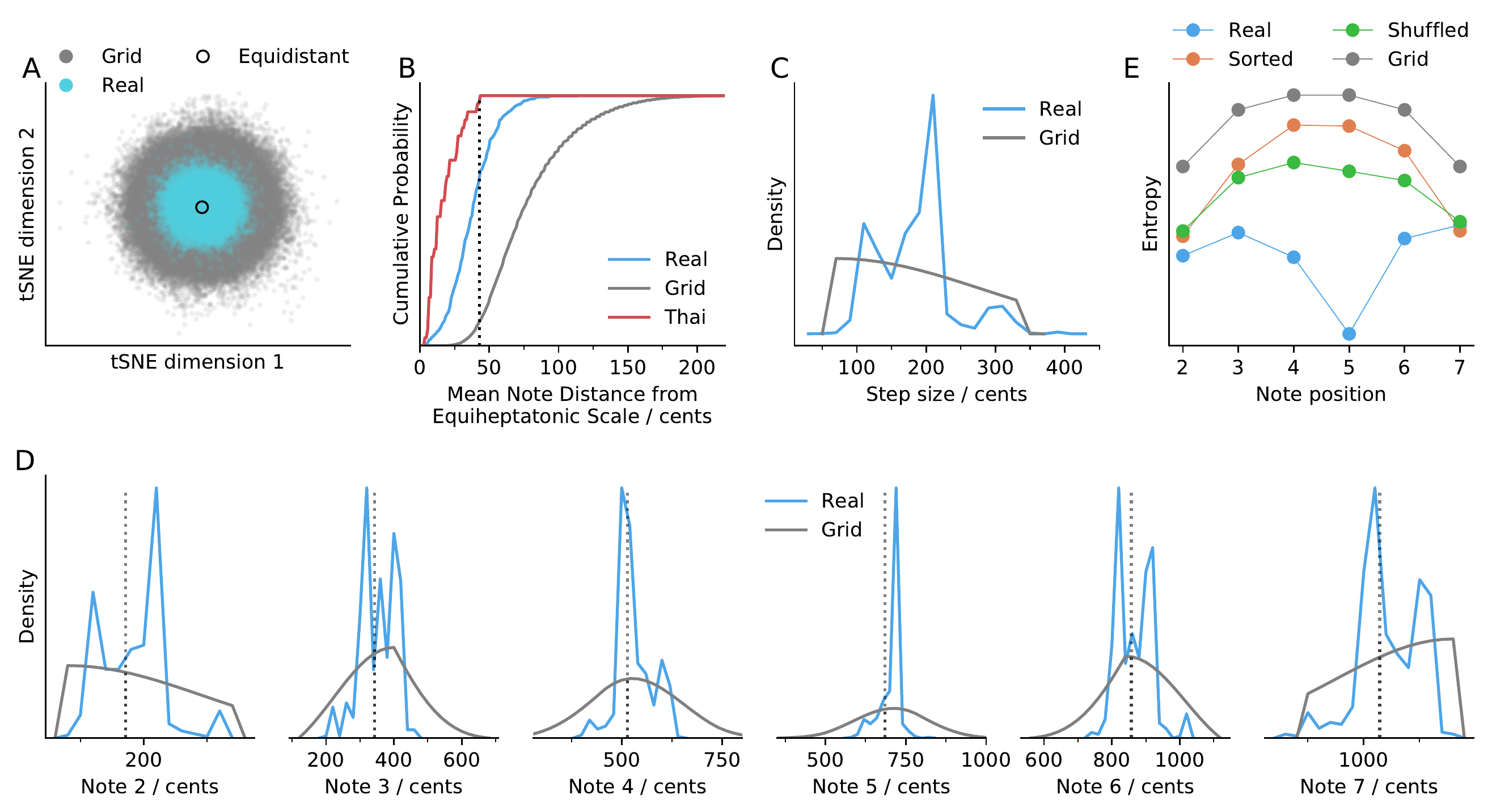}
  \caption{\label{fig:fig7}
  Comparison of real scales with all possible 7-note scales enumerated on a grid (\SI{20}{cents} resolution; with step sizes limited to \SIrange{60}{320}{cents}).
  A: Two-dimensional embedding of grid scales. Grid scales that correspond to real scales (notes are on average within \SI{10}{cents} of a real scale) are highlighted cyan; a black circle shows the equidistant scale.
  B: Mean note distance from 7-note scales and the equiheptatonic scale, for Thai scales, real scales, and all grid scales. 
  C: Step size histograms (bin size = \SI{20}{cents}) in real scales and grid scales.
  D: Scale note histograms (bin size = \SI{20}{cents}) for notes \numrange{2}{7} (no tonic and octave) for
  real scales and grid scles.
  E: Entropy of note distributions for: real scales; real scales, but rearranged with their steps arranged in order of size (Sorted); real scales, but rearranged with their steps in random order (Shuffled); grid scales.
  }
  \end{figure*}

\nsb{Statistics of Octave Scales.}
  After verifying that preferential use of octaves is indeed widespread, we proceed to studying octave scales;
  we infer octave scales by making some assumptions (\eg, octave equivalence; see section ``Inferring octave scales'').
  While in the previous section we exclusively examined measured scales, here we look at a mix of theory and measured scales. Data is shown for four samples of scales: (i) all theory scales, (ii) all measured scales, (iii) sub-sample controlling for society (`SocID', no more than 5 scales per society), and (iv) a sub-sample controlling for geographical region (`Region', no more than 10 scales per region).
  
  Most scales are found to have 7 or fewer notes (\fref{fig:fig5}A), in agreement with previous work.~\cite{savpn15} 6-note scales are remarkably rare across all sampling schemes.
  The predominance of step sizes of $\SI{\sim 200}{cents}$ (\fref{fig:fig3}B) is also seen in octave scales (\fref{fig:fig5}B). Consistent with existing literature,~\cite{pat10} most steps are between \SIrange{100}{400}{cents}, and this applies in all geographic regions studied (SI Fig. 5). The most common notes (\fref{fig:fig5}C), in all sub-samples, are exact matches for the significant intervals in \fref{fig:fig4}B (\SI{500}, \SI{700}{cents}). 
  The fact that the statistics of octave scales is consistent with the statistics of measured scales
  (and with previous work), to an extent, validates our methodology for extracting octave scales (SI section ``Inferring octave scales'').

  Theory scales show sharp peaks at intervals close to 12-TET intervals. In contrast, the distributions of steps and notes in measured scales are much more diffuse. This difference is expected, since theory scales consist of mathematically-exact, ideal intervals, while measured scales include natural sources prone to error and variation.   There is, however, correspondence between theory and measured scales in the rarely used notes (\SI{270}, \SI{450}, \SI{550}, \SI{650}, \SI{930}{cents}), which is even clearer when controlling for the number of notes (SI Fig. 8).
  Regardless of the number of notes in a scale, we find salient peaks at \SI{200}, \SI{500} and \SI{700}{cents} (SI Fig. 8). Overall, despite some differences between theory and measured scales, both sets are in agreement over the most distinct features: (1) step sizes
  are usually $\sim$\SI{200}{cents}, and between \SIrange{100}{400}{cents}, and (2) the most salient notes are at \SI{500} and \SI{700}{cents}, while the regions around these notes are avoided in scales.\\

\nsb{Variation across societies is comparable to variation within some societies.}
  To quantify variation in scales across societies, we use t-distribution stochastic neighbour embedding (tSNE) to map the scales onto a reduced two-dimensional representation.~\cite{vanjm08} This method can compare only scales with the same number of notes, so we show results separately for 7-note scales (\fref{fig:fig6}A), and 5-note scales (SI Fig. 9). We group scales into clusters (using the DBSCAN algorithm (eps = 2, minimum samples = 5),~\cite{pedjm11} and report note distributions and region distributions for the largest four clusters.
  The tSNE embedding is useful for visualizing diversity among high-dimensional objects,
  but note that tSNE dimensions are arbitrary. The main information in this plot is 
  the distances between points: distances in the 2d-embedding are non-linearly related to the real distances,
  such that 2d distance from one scale to all other scales will have a high rank-correlation with the real distances.

  The simplest and most salient feature is that scales tend to be almost \emph{equidistant} (\fref{fig:fig6}D). For example, in the largest cluster (\fref{fig:fig6}A-B; blue), notes are all within \SI{30}{cents} of the corresponding notes in the equiheptitonic scale. Although most examples (29 / 40) of this cluster (\fref{fig:fig6}D) come from three countries (Thailand, Guinea, Malawi), in total they are found in \num{13} countries (\eg, Colombia, Georgia, Zimbabwe, Indonesia). Surrounding this cluster (\fref{fig:fig6}A), we can find examples of theory scales: Western diatonic modes, and North Indian thaat (\eg, green cluster).
  In general, we see that most equidistant scales are measured scales, but there is also a lot of overlap between theory and measured scales (\fref{fig:fig6}E). Furthermore, when clustering societies by scale similarity, we find two main clusters -- one dominated by societies with theory scales, and one dominated by societies with equidistant scales (SI Fig. 10-11). Examples of scales that are furthest from equiheptatonic include the Gamelan pelog scale (\fref{fig:fig6}A-B; yellow cluster), and the Carnatic mela salagam (\fref{fig:fig6}F), which both have many small step sizes in a row. 
  
  Surprisingly, there seems to be little variation overall, such that multiple societies
  that use theory scales exhibit levels of within-society variability comparable to
  the total variability (\fref{fig:fig6}F). In general, we find that geographical regions
  tend to contain overlapping subsets of scales (SI Fig. 12A), and distances between scales
  within-regions are of similar magnitudes to distances between scales between-regions (SI Fig. 12B).
  Even Thai scales, which are often cited as exclusively being in equiheptatonic tuning \cite{mor76,attpi04,wis15,moric18} -- although this is disputed \cite{garan15} -- exhibit substantial within-society variability (\fref{fig:fig6}F). Overall, there appears to be less variation in 7-note scales than might have been expected, and a similar analysis of 5-note scales reveals similar results (SI Fig. 8), which suggests that they share the same organizing principles.\\

\nsb{Statistical analysis shows that scales tend to be equidistant.}
  To put the diversity of scales in a broader context, one may consider a hypothetical universe of possible scales by enumerating them on a grid (\textit{grid scales}).
  To this end, we take \SI{20}{cents} as a basic grid resolution, and enumerate all possible unique scales with step sizes within \SIrange{60}{320}{cents};
  this limitation on step size already reduces the number of possible scales (at \SI{20}{cents} resolution) by \SI{98}{\%}.
  To conveniently compare grid and real scales, we examine their 2-dimensional embedding. Strikingly, those grid scales that correspond to real scales (\ie, within \SI{\pm 10}{cents} of a real scale) are clustered around the equiheptitonic scale (\fref{fig:fig7}A, cyan).
  For comparison, Thai scales are found to be as far as \SI{43}{cents} from equiheptatonic. We find that \SI{70}{\%} of real scales are within this boundary, compared to only \SI{10}{\%} of grid scales (\fref{fig:fig7}B).
  Alternative definitions of equidistance based on step sizes rather than scale notes
  also support the finding that scales tend to be close to equidistant (SI Fig. 11).

  To gain further insight into the striking (near-) equidistance of scales,
  we look at the note distributions for notes \numrange{2}{7} (\fref{fig:fig7}D)
  for real scales and grid scales. The grid scales show the broadest note distributions for
  the two notes (4 and 5) furthest from the fixed ends (tonic, octave), resulting in higher entropy (\fref{fig:fig7}E).
  This is expected, since summing two random distributions should result in a higher variance;
  \eg, note 4 is the sum of three steps drawn from \fref{fig:fig7}C.
  In contrast, distributions of notes 4 and 5 in real scales are most predictable (lowest entropy),
  once again illustrating the significance of intervals of size \SI{500} and \SI{700}{cents}.

To examine the possibility that the difference between real scales and grid scales is mainly due to differences in their step size distributions (\fref{fig:fig7}C), we analyse two other sets of note distributions. We first look at the note distributions that ought to be farthest from equidistant scales: we rearrange the steps in each real scale so that they are ordered low-to-high, and consider scales starting from every position. This results in note distributions with entropy similar to grid scales (\fref{fig:fig7}E, Sorted).
  We then look at the note distributions of alternative versions of real scales obtained by shuffling the step sizes in scales. Again, we find that the entropy is highest at notes 4 and 5 (\fref{fig:fig7}E, Shuffled),in contrast to real scales.
  The main difference between the real scales and grid scales appears to be that step sizes in real scales
  are well-mixed -- small steps are found adjacent to large steps rather than small steps (and vice versa).

\section*{\sb{Discussion}}

\nsb{How diverse are scales?}
  Despite some differences, scales across cultures are \emph{remarkably similar}. For example, the set of Carnatic scales alone is almost as varied as the total set of scales (\fref{fig:fig6}F). In particular, most scales observed in cultures -- when compared to the universe of possible scales -- are close to 5- and 7-note equidistant scales (\fref{fig:fig7}, SI Fig. 12). 
  Many authors have reported that equidistant scales are rare.~\cite{pat10,mcdmp05,bal10,bropm13,savpn15,rosms17}  Yet, we find that they are more prevalent than expected by chance. This discrepancy may be due to the lack of a robust definition for equidistant scales -- for example, the only previous statistical study of prevalence of equidistance does not explicitly account for natural variation in intonation.~\cite{savpn15}
  It is not clear how much deviation from perfect equidistance renders a scale perceptually non-equidistant, or how variability in pitch affects perception of equidistance. As an illustration of the difficulties, consider the following example from Georgian singing: the step sizes in this scale are close to equidistant (\SIrange{163}{202}{cents}) when viewing the melodic pitch histogram. But the intervals between notes in the melody are much less exact (\SIrange{90}{240}{cents});~\cite{sch20} despite being statistically equidistant, it is not clear whether this scale would be perceived as equidistant. Perception of equidistance may also vary with culture and training, in which case it may be better to avoid binary measures of equidistance, and instead to construct a perceptually-relevant, continuous measure.

  The most salient difference between societies is whether their data consisted of theory scales or measured scales (SI Fig. 9-10), considering that theory scales appear to be less equidistant than measured scales. One hypothetical explanation for the predominance of non-equidistant scales in societies using theory scales: the process of creating new scales by combining simple-integer-ratio intervals will inevitably result in more non-equidistant scales because only one combination of step sizes can result in an equidistant scale. 
  We wonder whether this societal difference would persist if we were to only compare measured scales from societies, since theory scales will certainly  exhibit intonation variability when performed. In the literature, Western scholars often discuss variation in intonation in societies that lack mathematical musical theory as being an intentional form of expression,~\cite{rahyi78,coobp82,vetet89,permq94,set05,garan15}   whereas studies of classical musics typically investigate to which theoretical tuning system the musicians conform.~\cite{hagst80,seris11,damjv20,sch20}   Ultimately, it is difficult to intuit the differences between societies, since there are only a few cross-cultural studies of interval discrimination.~\cite{arocm93,permp96} Here, we performed a preliminary study on intonation variability (SI Fig. 13-14), finding comparable levels of variation (with the possible exception of the pelog scale) in Gamelan orchestras, Thai xylophones, Turkish ney,~\cite{tan11} Georgian singing,~\cite{mzhmu20} and a Belgian carillon.~\cite{sch17} But ultimately, to understand the differences between societies that use theory scales and those that do not,   we will need cross-cultural perceptual experiments, and direct measurements of scales from recordings.

  How did far-away societies come to use such similar scales? --
  Using a cultural-evolutionary framework,~\cite{boy88} we can describe this process as a combination of
  cultural diffusion, and convergent evolution due to common factors biasing the use of scales across cultures.
  Diffusion can certainly account for some similarities, as some societies have documented shared history:
  \eg, societies with theory scales in the Middle East, or East Asia; court music in Thailand, Laos and Cambodia.~\cite{chu15}
  Some have suggested that cultural diffusion can be inferred based on two cultures using similar
  equiheptatonic scales,~\cite{jonam60,min90} but evidence seems circumstantial
  given how widespread equiheptatonic scales are.
  Instruments, on the other hand, are a richer source of information than scales, so organology studies are
  a much more direct way of inferring cultural diffusion.~\cite{kiras66,bleaa14,agueh20,aguhs21} 
   While there is undoubtedly some transmission of information across cultures, 
   we reiterate that:
   within-region scale diversity is comparable to between-region diversity (SI Fig. 12B),
   step intervals consistently show approximate limits of \SIrange{100}{400}{cents} across regions (SI Fig. 5),
   and scales are (against chance odds) overwhelmingly close to equidistance in all regions (SI Fig. 12A).
  Taken together, these facts point towards some non-negligible degree of convergent evolution due to shared biases. \\

\nsb{How do scales evolve over time?}
  Scales can change, or persist, in a variety of ways. On a short time-scale, vocal (or other non-fixed-pitch instrument) scales are inherently stochastic, due to a lack of precision in motor control.~\cite{hutje12} On a longer time-scale, vocal scales reside in memory, which introduces another mechanism of change. Unlike vocal scales, instrument tunings physically persist through time, but are still affected by multiple factors: environment (temperature, humidity), material (wood, metal, animal organs), and physical force.~\cite{kubam80} To illustrate this point, we checked examples of repeated tunings
  of the same instruments across a specific time frame:  Gamelan orchestra (metal idiophone), with standard deviation of notes, $\sigma=$ \SI{8}{cents} (slendro) and $\sigma=$ \SI{13}{cents} (pelog) cents over about 25 years;~\cite{sur72} Angolan likembe (plucked metal idiophone), where $\sigma=$ \SI{18}{cents} 
  over a few weeks;~\cite{kubam80} Gambian kora (chordophone), with $\sigma=$ \SI{27}{cents} over one week.~\cite{anija82} Technology offers more robust ways of keeping scales stable over time: monochords (ancient Greece) and pitch pipes (ancient China) enable reliable   tuning by fourths and fifths.~\cite{rec18}
  Likewise, the perceptual phenomenon of tonal fusion may have enabled stable scales by providing knowledge of perceptual anchors (octaves and fifths) that can be reliably passed through generations.
  Recently, it seems to us that stability in scales has been reinforced though global tuning standards and inventions such as fixed pitch instruments and electric tuners. Thus, we can see that scales change in many ways, but it is possible   for the rate of change to decrease through the use of technology and musical theory.

  We can imagine some possible mechanisms of scale evolution, drawing analogies
  with biological evolution: We can think of a single scale as a gene,
  and a set of scales used by a population of people as a genome.
  Changes to the scale notes is akin to mutating a single nucleotide.
  Adding or removing notes from a scale is like insertions/deletions of nucleotides.
  Scales can be copied through interactions between populations; genes can be shared through
  horizontal transmission.
  The Western diatonic modes can be constructed using the same step intervals, but changing the tonic position;
  the same process occurs in protein sequences (circular permutants), likely through gene duplication and homologous recombination.

  The analogy starts to break when you consider that scales can be invented, which is more akin to designing genomes. 
  There are recent examples of invention of microtonal tunings in Western music,~\cite{par50,set05,rec18}
  and many theory scales bear the hallmarks of design: Greek modes are all circular permutants,
  based on simple integer ratios;~\cite{croja63} the Carnatic melakarta result from combinatorial enumeration of a set of intervals, constrained by a set of rules.~\cite{vij07} Similarly, other cultures that have a long history of mathematical scholarship \cite{go16} use overlapping sets of scales based on theoretical divisions of the  octave.~\cite{maram93,kat15,far04,sig77,te70,vonja92}
  Typically, though, the theory scales that still survive are similar to the meaured scales,
  so it is likely that designed scales are constrained by similar selection criteria
  that applies to non-designed scales.

  We propose that a detailed mechanistic model of scale evolution is possible:
  The mechanisms of spontaneous change need to be studied at the resolution of a single mutation step (through, \eg, transmission chain experiments).\cite{angcb23}
   The relative importance of
  the different mechanisms could be investigated with sufficient long-term data
  (\eg, same individual/group/culture, over a period of months/years).
  Such a model could then be used for agent-based modelling to study the role
  of horizontal transmission, and cultural-evolutionary biases.
  \\

\nsb{How are scales selected?}
  We propose three types of selection pressure:
  (i) cultural-evolutionary biases are based on how many people, or which people in one's group are using the scale;~\cite{crepn17} (ii) cognitive biases depend on how pitch is perceived by humans;
  and (iii) production biases depend on what is easy or difficult to sing, and physical constraints on instruments.

  We can think of three examples of cultural-evolutionary biases that
  may apply to music: conformity, novelty and presitge biases.
  A bias towards conformity results in the most populous scale type
  being increasingly successful.
  In general, humans tend to synchronize~\cite{savbb21,mehbb21} when playing music (same meter, tonality),
  and over a longer time period, there is evidence of shifts towards the use of 12-TET.~\cite{seris11,moeis09}
  A conformity bias could arise due to other reasons, but we can probably rule out the difficulty of learning
  scales -- studies on novel scale systems have shown that people rapidly learn their statistics.~\cite{loump10,rohcs17}
 Novelty biases could be can describe choosing novel microtonal tunings as a means of expression.~\cite{par50,vetet89} An example of a prestige bias is where tunings are copied from players/instruments that are acknowledged as good.~\cite{sur72,copef18}

  Theories of scale selection are commonly based on cognitive biases,
  of which we will give a brief summary:
  liking harmonicity,~\cite{hurmp94,gilpo09} disliking sensory dissonance~\cite{auclm08,bersa19}
  (or liking sensory dissonance \cite{ambji09}), tonal fusion,~\cite{stu90} neurodynamic theory,~\cite{larmp16}
  transmissibility,~\cite{stu12,dowja88,pfojc17,phi22,angcb23} memory,~\cite{dowpr78}
  and theories based on mathematical properties of scales.~\cite{clomt99,harjm20,balcm80,treje99,pelpn21,vermp17}

  Production constraints can apply to both singing and instruments.~\cite{tiepn11} When singing, large intervals cost more energy to produce than smaller ones, while small intervals are difficult to produce reliably due to limits on motor control.~\cite{sunjv93}
  Instruments are less constrained in this way, but still have physical limits to the number of notes, and interval range.
  On the other hand, instruments are constrained in how reliably they can be re-tuned to the same scale.
 There is a long history~\cite{croja63,wesml94} behind the practice of tuning using harmonic intervals,~\cite{vanam80,anija82,mil85,traam91,kus10}
 and reports exist of tuning according to the step sizes,~\cite{wacna50,kubam80,zeme81} tuning instruments visually,~\cite{zemet79,mil85,olg98,copef18} and copying a reference instrument.~\cite{sur72,copef18}
  \\

\nsb{How can we study the evolution of scales?}
  Tracing the evolutionary history of scales is challenging. Scales change at rates that depend on the instruments and technology, and new scales can be invented from scratch. The evolution is potentially driven by numerous selection pressures that vary in strength across societies. Nonetheless, we propose some approaches that seem feasible.

  We can try to track evolutionary trajectories of scales using historical data. By restricting the scope to a single society, one can make simplifying assumptions by using ethnographic accounts to inform models. For example, about Gamelan orchestras we know that they are typically tuned in reference to another orchestra,~\cite{vetet89} and instrument intonation changes at a steady rate (due to similar materials and environment). This can be described as an evolving network of Gamelan orchestras with edges between orchestras that influence each other. Gamelan tunings are also extensively documented in the literature, and additional tunings can be inferred from recordings.

  Another approach to infer the evolutionary trajectories or selection pressures is by analyzing a population of scales. This approach requires appropriate mathematical models where multiple selection pressures can be considered in tandem. Ultimately, many models may have convergent predictions, which means that additional experiments are needed to distinguish the relative importance of different selection pressures. One limitation of this approach is the dependence on the sample of scales studied, and it is hard to construct \textit{a priori} a representative sample of scales. In this study we control for society, but other criteria are possible. 
  Does it matter if societies have different population sizes? How does one deal with differences in within-society variation in scales? Should we take into account frequency of scale use within a society?~\cite{gelgu02,chu15}
  These questions will surely unravel as more data becomes available for hypothesis testing. \\

\nsb{Possible bias in the scale database.}
  Relying on data collected by a limited number of ethnomusicologists, the database at this stage has sparse geographic coverage (\fref{fig:world}). Some have suggested that ethnomusicologists have a bias towards reporting findings that are considered `interesting', thus exaggerating diversity.~\cite{net10} Certain musical traditions, such as Gamelan and Thai, were very popular research topics, so they are over-represented. In contrast, there are very few quantitative measurements of scales or instruments with fewer than 5 notes, despite these being reported in many sources. Additionaly, vocal scales (taken from recordings of singing) are scarce in the database, yet the voice is probably the most important instrument from an evolutionary point of view.~\cite{pfojc17}

  Unfortunately, in some rare instances there seem to be statistical irregularities in the reporting of tunings: \citet{sur72} note that Jaap Kunst \cite{kun49} reported gamelan tunings (not included in this database) where all the higher notes were exactly an octave above the lower ones, and that this is extremely unlikely; in a study of prehistoric bone flutes where the tunings are given to an accuracy of \SI{1}{cent}, one flute is recorded as having a series of equal tempered intervals: $[200, 200, 200, 300]$.~\cite{zhaa04} Ultimately, these biases do not void the conclusions, but the results shown here will certainly need to be updated as more data becomes available.

\nsb{Limitations to studying scale evolution.}
  We may be witnessing a decline in diversity of scales due to the
  converging forces of globalization and technological change.~\cite{cow09} There is already evidence of homogenization, via the adoption of 12-TET.~\cite{moeis09,seris11} Thus, to understand how scales evolve, we must look to the past, but therein lies a different problem: the older the instrument, the less certain we are about how they were played. For prehistoric artefacts, we cannot be sure whether they (or their reconstructions) faithfully resemble the instrument in its original condition. So far, the the only instruments that remain sufficiently intact to play are aerophones,~\cite{haee79,yuapp87,zhaa04} which can produce many different scales depending on how they are played.~\cite{olg98,san05,rawwm07,shias15} Thus, we believe that the best source of scales is in ethnographic recordings spanning the past century.~\cite{woopo22} It is therfore imperative that methods be developed that can faithfully infer scales from large samples of songs. Algorithms must be developed to handle low-quality recordings,~\cite{gfeis20} background noise, instrument / singing segmentation,~\cite{marap19} polyphonic stream segmentation,~\cite{benis19} note segmentation,~\cite{maupf15} and tonal drift.~\cite{mauja14}

\section*{\sb{Conclusion}}

  Scales are a cornerstone of music across the world, upon which endless combinations of melodies can be generated. Surprisingly, despite a wealth of ethnomusicological research on the subject, we lacked a comprehensive, diverse synthesis of scales of the world. Here we remedy this issue, with a focus on quantitative data that will enable detailed statistical analyses about how scales evolve. Our own preliminary analyses have lent quantitative and qualitative support for the widespread (but not necessarily universal) use of the octave in some special capacity. Despite the rich diversity of scales, when put in context of how many scales are possible, what stands out in our analysis is how remarkably similar they are across the globe.
  Altogether, this work presents a treatise on the evolution of scales, and proposes promising avenues for future research.\\

{\smaller[0.5]
\nsb{Author Contributions.~} JM and SP were involved in creating the database; JM performed statistical analyses; TT supervised the project and edited the manuscript; JM wrote the manuscript with input from TT, SP.}\\

{\smaller[0.5]
\nsb{Data and Code Availability.~}
  DaMuSc (Database of Musical Scales) is available at \href{https://github.com/jomimc/DaMuSc}{https://github.com/jomimc/DaMuSc}.
  An archived version of DaMuSc, along with data needed to recreate figures is available at Zenodo (\href{http://doi.org/10.5281/zenodo.7250281}{http://doi.org/10.5281/zenodo.7250281}).
  Analysis code and code for creating figures is available at \href{https://github.com/jomimc/DaMuSc\_Paper\_Code}{https://github.com/jomimc/DaMuSc\_Paper\_Code}.
}\\

{\smaller[0.5]
\nsb{Acknowledgements.~} We thank Olga Velichkina, Frank Scherbaum, Malinda McPherson for discussions and comments on the manuscript. We thank Polina Proutskova, Elizabeth Philips, Steven Brown, Patrick Savage, Kaustuv Kanti Ganguli, John Garzoli and Neil McLachlan for discussions.}

\bibliographystyle{unsrtnat}
\bibliography{world_scale}

\end{document}


\maketitle

\begin{figure}[t!]  \centering
  \includegraphics[width=0.85\linewidth]{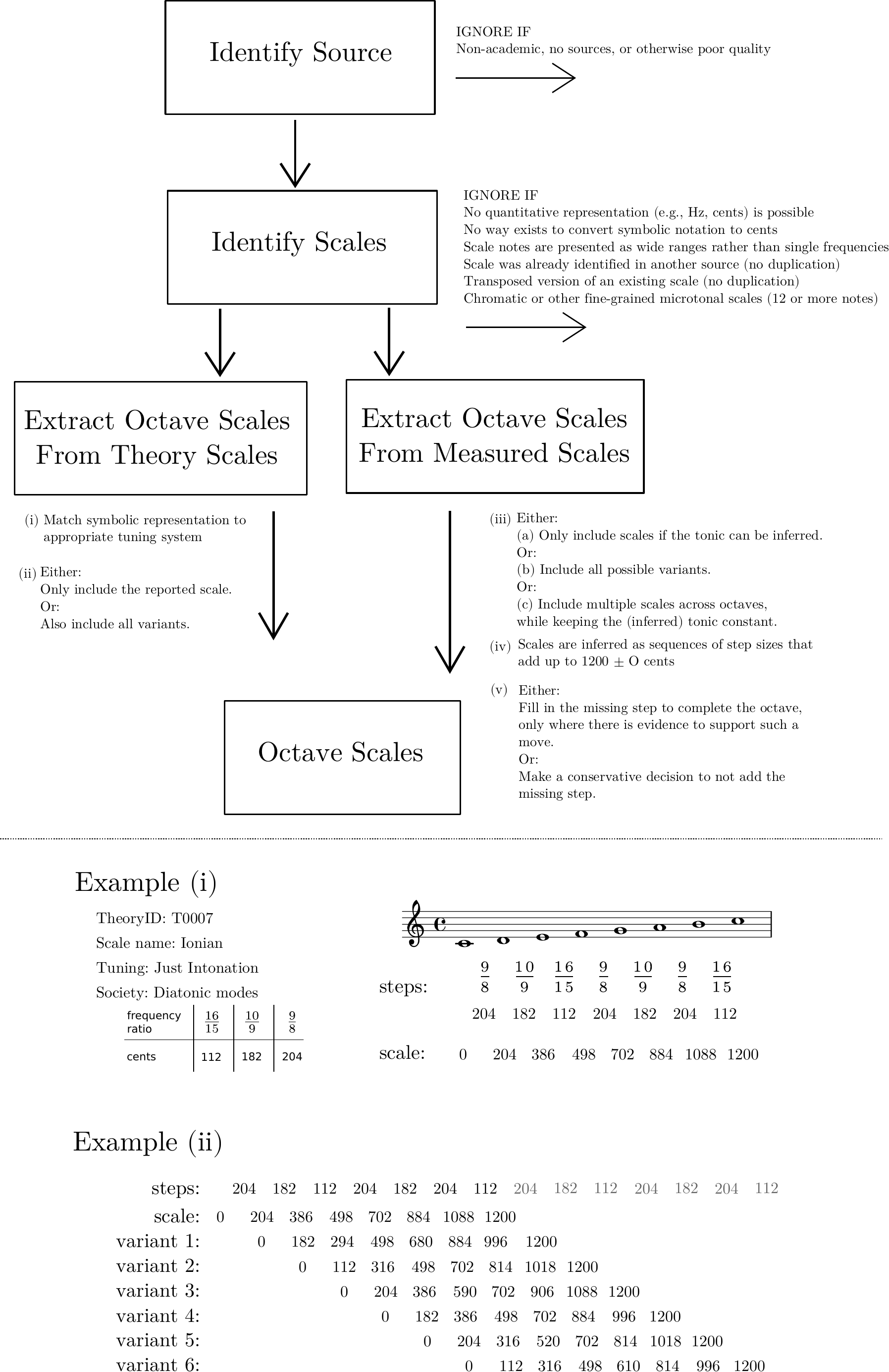}
  \caption{\label{fig:si1}
  Top: Schematic for construction of scale database. Appropriate sources are first
  identified, and out of these we identify scales that have unambiguous quantitative
  representations. This results in a set of `theory scales', and a set of `measured scales'.
  From these two sets, we create a set of `octave scales', according to a set
  of five (i -- v) choices.
  Bottom: Examples of how choices (i) and (ii) are implemented.
  (i) Most theory scales are represented symbolically, and we require some
  kind of code to convert the symbolic representation into a quantitative representation.
  (ii) For purposes of analysis, if one wants to ignore the concept of
  tonality (here referring to the idea that positions in a scale are
  not equivalent, and that scales start at the tonic), one can
  include all possible variants that start on different positions.\\
  ...
  }
  \end{figure}

\begin{figure}[t!]  \centering
  \includegraphics[width=0.90\linewidth]{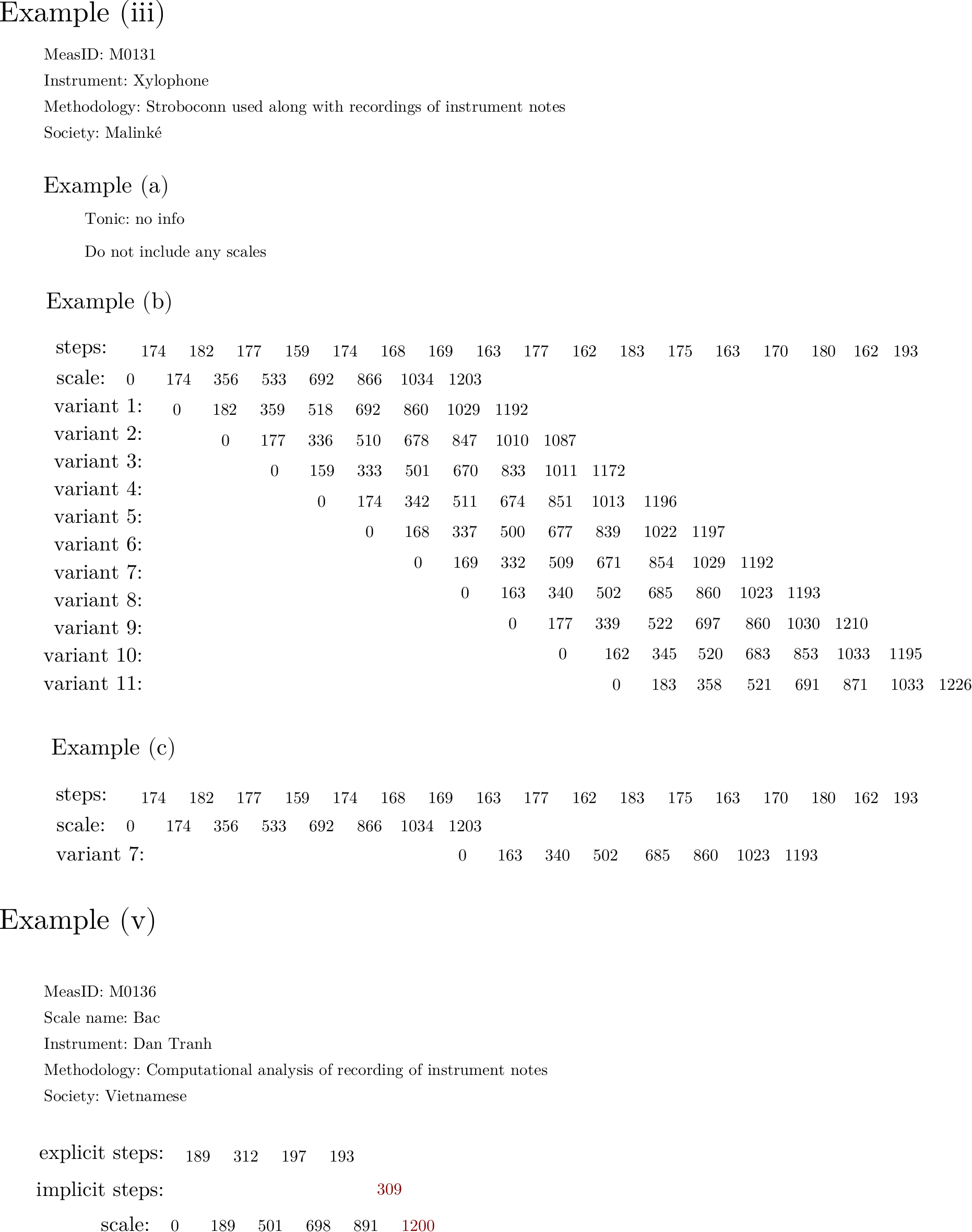}
  \caption{\label{fig:si2}
  continuing from \fref{fig:si1}... \\
  (iii) sources that contain `measured scales' rarely explicitly include detailed information
  about tonality, let alone which note may be considered a starting note (`tonic').
  Thus, one can choose to: (a) Only include scales if there is evidence for which note(s) the
  performer considers a tonic.
  (b) Include all possible variants where the final note sums to $1200 \pm \mathcal{O}$ cents.
  (c) Identify one potential `tonic', and only include scales that start on that
  note, or on notes that are related to the `tonic' by an octave relation.
  (iv) The value of $\mathcal{O}$ needs to be specified; we choose $\mathcal{O}=50$ cents.
  (v) Some sources indicate that a full octave is used, and for reasons of concision they
  do not report the final interval leading to the octave. In this case, one can choose to
  include the final interval (shown in red) or not.
  }
  \end{figure}

\nsb{Inferring octave scales.}
  The database exists in two forms: the raw data (382 theory scales, and 434 measured scales), and a set of octave scales that is generated procedurally from the raw data according to five choices. For a complete workflow from source to database, including examples, see \fref{fig:si1} and \fref{fig:si2}. The five choices are:

  \textbf{(i) Theory scale intonation:}  Theory scales are typically written in symbolic notation (\eg, solf\`{e}ge, swara), which can then be converted to numerical sequences by matching the symbols to a tuning system (\eg, 12-tone equal temperament, 12-TET). For each theory scale in the database, we provide a set of tuning schemes that were plausibly used, so a single theory scale may have several versions in different tunings.

  \textbf{(ii) Theory scale variants:} The tonic is the first note in all theory scales. However, one may also ignore tonality and include all variants from all theory scales by taking different notes as starting points, thereby forming circular permutations of the original scale. This would create many duplicates since theory scales tend to contain many scales that are
  variants of each other.

  \textbf{(iii) Measured scale variants:} Tonality is rarely clear in sources of measured scales, and therefore it is impossible to uniquely identify the order of notes as intended by the performer. For these scales, one can: (i)  include only those where the tonic is known; (ii) include all plausible scales (all variants); (iii) or include more than one scale if there are multiple octaves, without including variants. We define a plausible scale on an instrument tuning as a sequence of notes that span an octave.

  \textbf{(iv) Measured scale octave error tolerance:} In practice, measured intervals exhibit random fluctuations, so we accept scales if the intervals sum to $1200 \pm \mathcal{O}$ cents, where $\mathcal{O}$ is the deviation tolerance.

  \textbf{(v) Measured scale missing octave addition:} In some cases, sources indicated that the octave was used in performances, but reported the scales without the octave. For these cases, we can choose to include the missing interval that is needed to complete the octave (affects \num{41} out of \num{434} samples). This happens when, for example, measured scales inferred from recordings are collapsed onto a single octave; in this case, we assume that the final interval was omitted simply because it is redundant information.

\begin{figure}[t!]  \centering
  \includegraphics[width=0.75\linewidth]{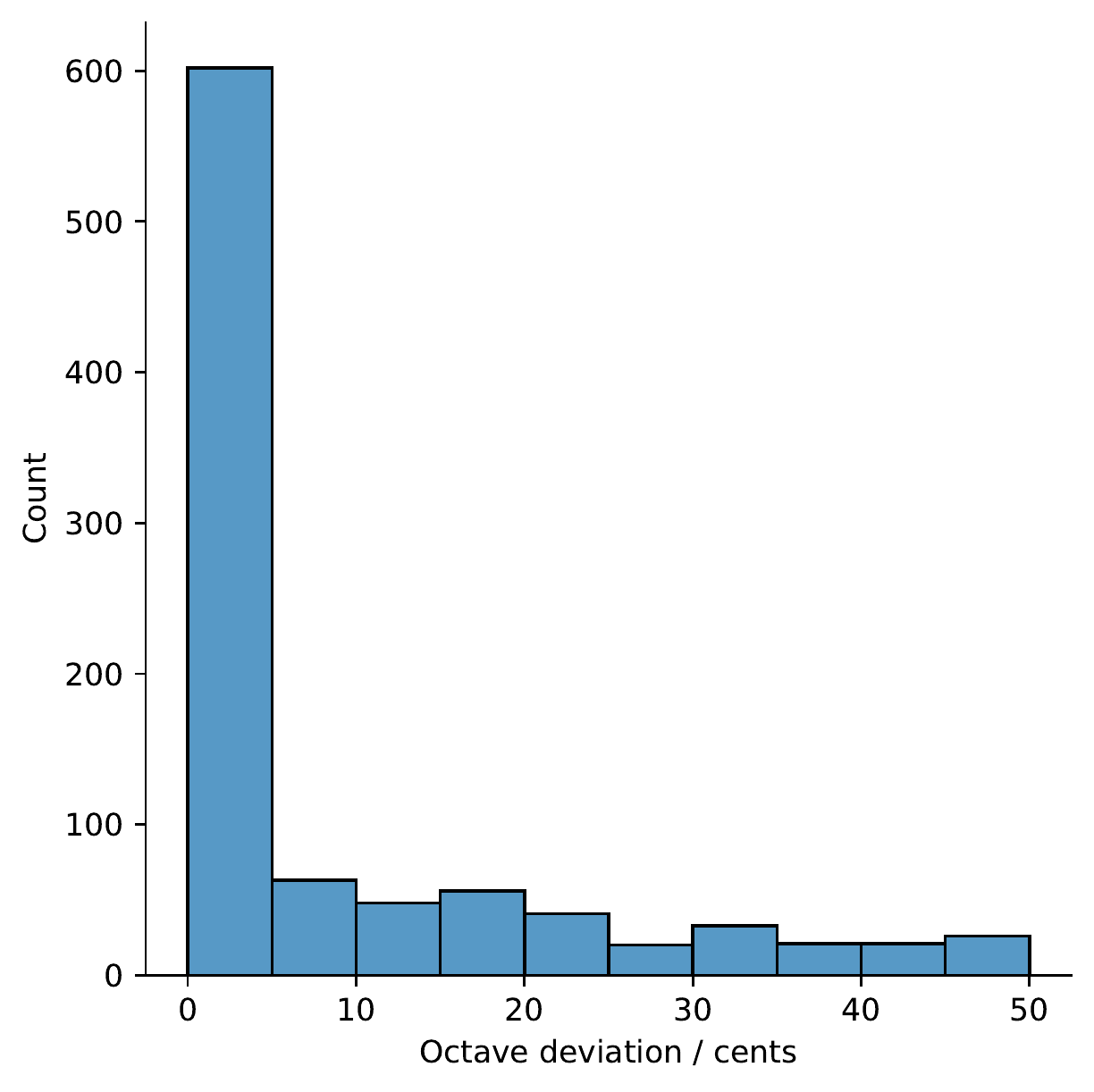}
  \caption{\label{fig:si3}
  Distribution of deviations of final notes from the octave for all octave scales ($\mathcal{O}=50$).
  }
  \end{figure}

  Different sets of choices introduce variations in the database.
  In the main manuscript we report statistics for the database created according to the following choices:
  (i) We match theory scales from each musical tradition to a set of tuning systems given in SI Table 2. 
  (ii) We do not include all possible variants of theory scales. 
  (iii) We do not exclude scales inferred from instrument tunings if we do not know the tonic, but we do not include all possible variants.
  (iv) We use a tolerance of $\mathcal{O} = 50$, noting that \SI{73}{\%} of extracted scales are within \SI{10}{cents} of the octave (\fref{fig:si3}).
  (v) We include scales that require an extra interval to complete the octave. 
  In total, this results in \num{896} scales (\num{434} theory scales, \num{384} scales from instrument tunings, and \num{78} scales from song recordings). The theory scales span \num{6} regions, while the   measured scales span \num{8} regions, and \num{46} countries (main manuscript Fig. 2).

  \clearpage

  \clearpage

  \clearpage

  \clearpage

  \begin{figure}[t!]  \centering
  \includegraphics[width=0.95\linewidth]{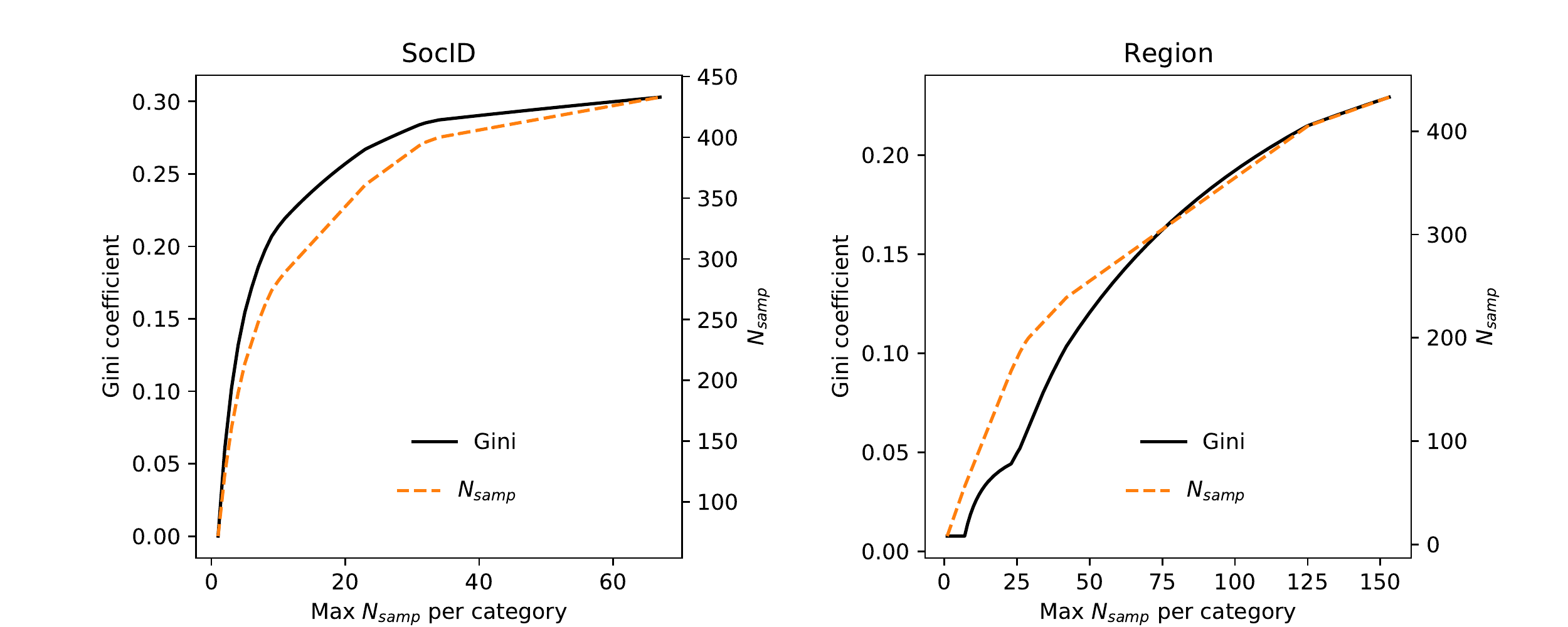}
  \caption{\label{fig:si4}
  Total number of scales $N_{samp}$, and Gini coefficient as a function
  of the maximum number of scales allowed per category: SocID, left; Region, right.
  }
  \end{figure}

  \clearpage

  \begin{figure}[t!]  \centering
  \includegraphics[width=0.95\linewidth]{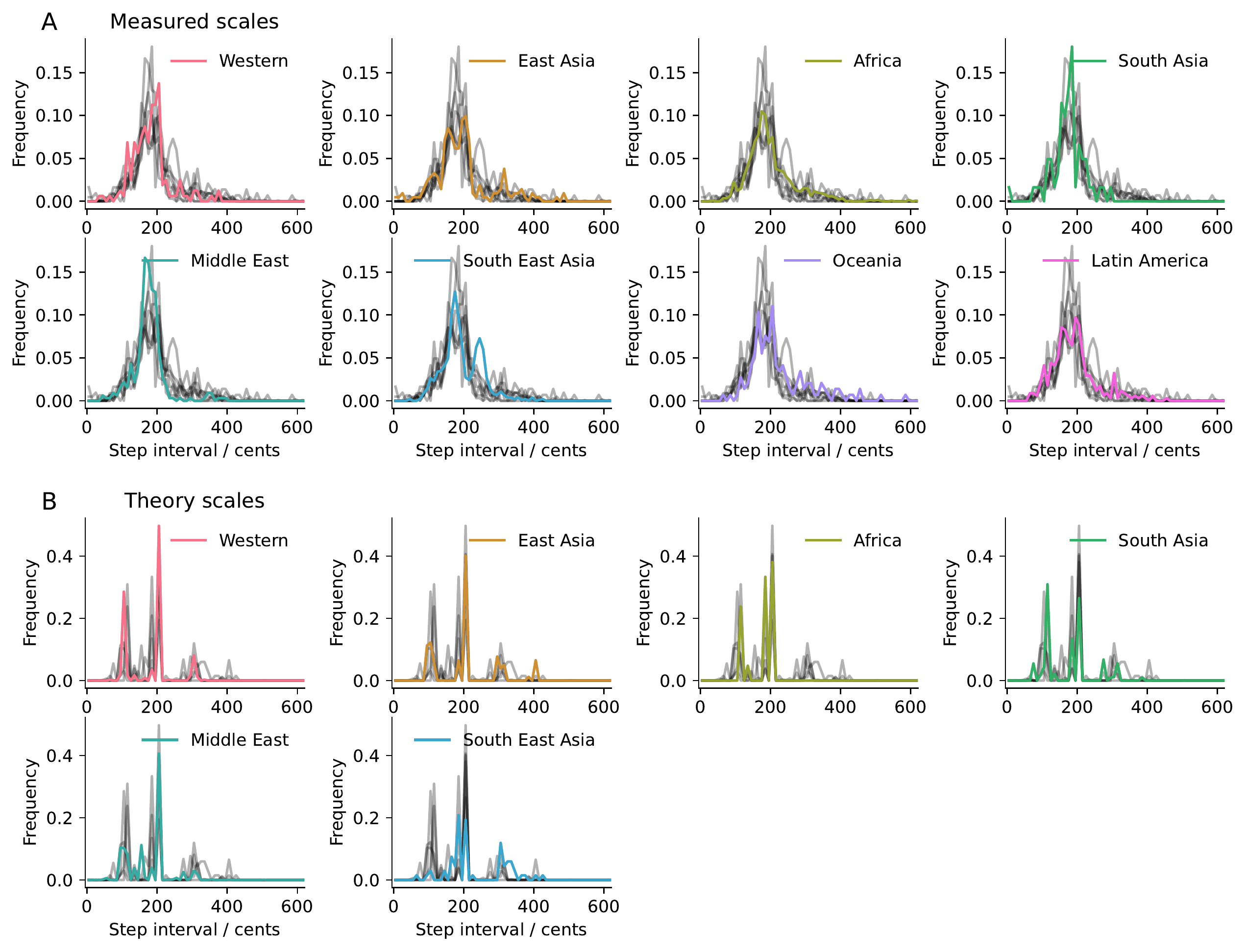}
  \caption{\label{fig:si4}
  Step interval distributions for Measured scales (A) and Theory Scales (B), from
  different geographical regions.
  In each panel, the distribution for one region is highlighted, and the distributions  
  for the other regions are shown in grey. There are no Theory scales in the database from
  Oceania or Latin America (B).
  }
  \end{figure}

  \clearpage

  \begin{figure}[t!]  \centering
  \includegraphics[width=0.95\linewidth]{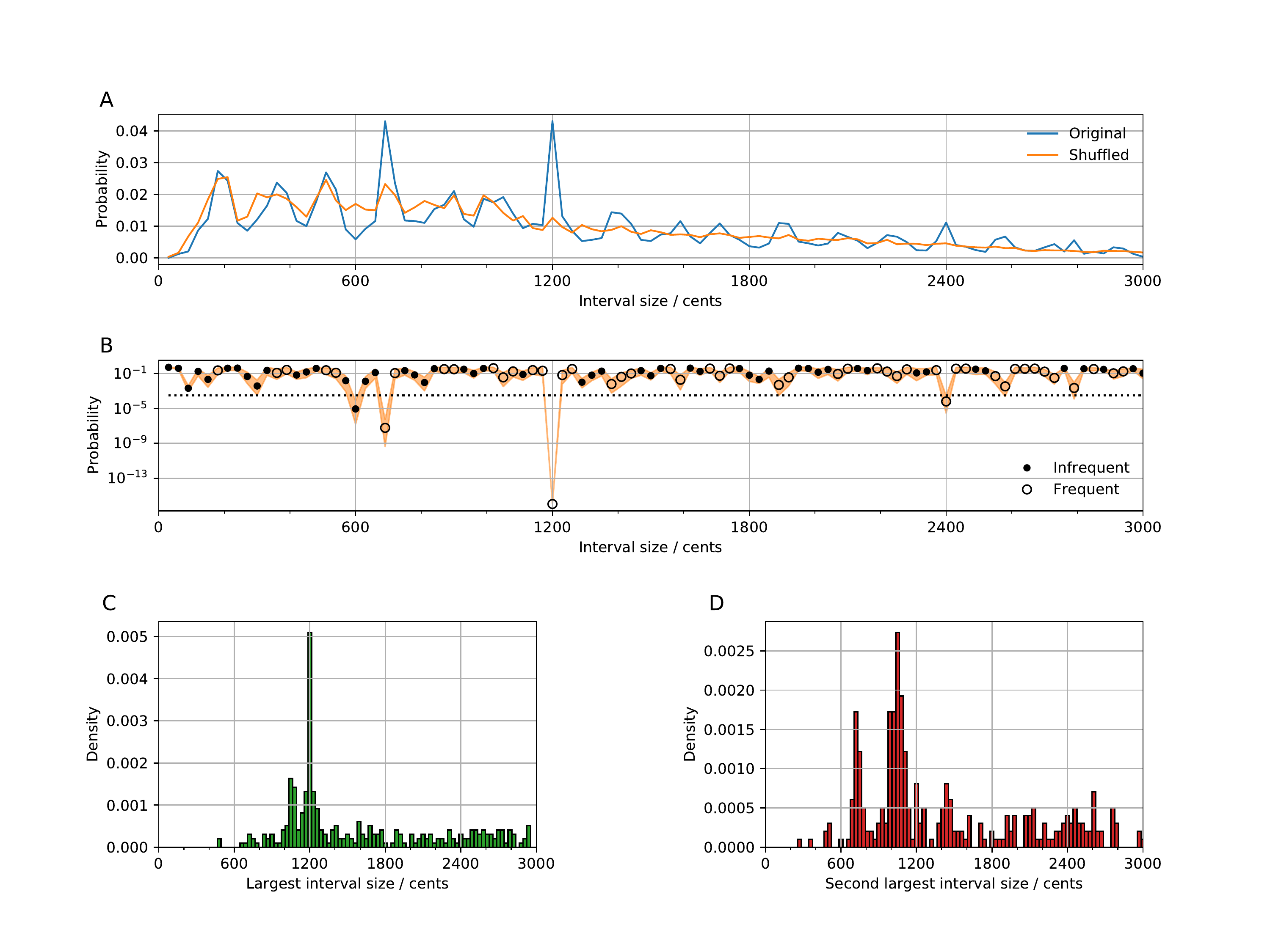}
  \caption{\label{fig:si5}
  A: Distribution of notes in measured scales (blue; histogram bin size = 30 cents).
  Distribution of notes obtained via alternative sampling:
  shuffling the step sizes within scales (orange), for those scales that are far
  from equidistant (difference between minimum and maximum step sizes is greater than \SI{100}{cents}).
  The x-axis is truncated at \SI{3000}{cents} for clarity.
  B: Probability that the counts observed in the original data were generated by
  the (far-from-equidistant) shuffled scales distribution. \\
  The peak at \SI{700}{cents} is lower than in the main text Fig. 3D.
  This indicates that for scales that are far from equidistant, the step sizes
  are arranged so that fifths appear more frequently than chance. \\ \\
  C: Distribution of the largest interval size for all scales.
  D: Distribution of the second largest interval size for all scales. \\
  When we create new samples by shuffling without replacement, we do not count the final
  note since this will always be the same. We considered that if there were many equidistant
  scales, they would result in octave intervals appearing by chance. However, this does
  not happen as the second largest interval size falls short of an octave
  in many of these equidistant scales.
  }
  \end{figure}

  \clearpage

  \begin{figure}[t!]  \centering
  \includegraphics[width=0.75\linewidth]{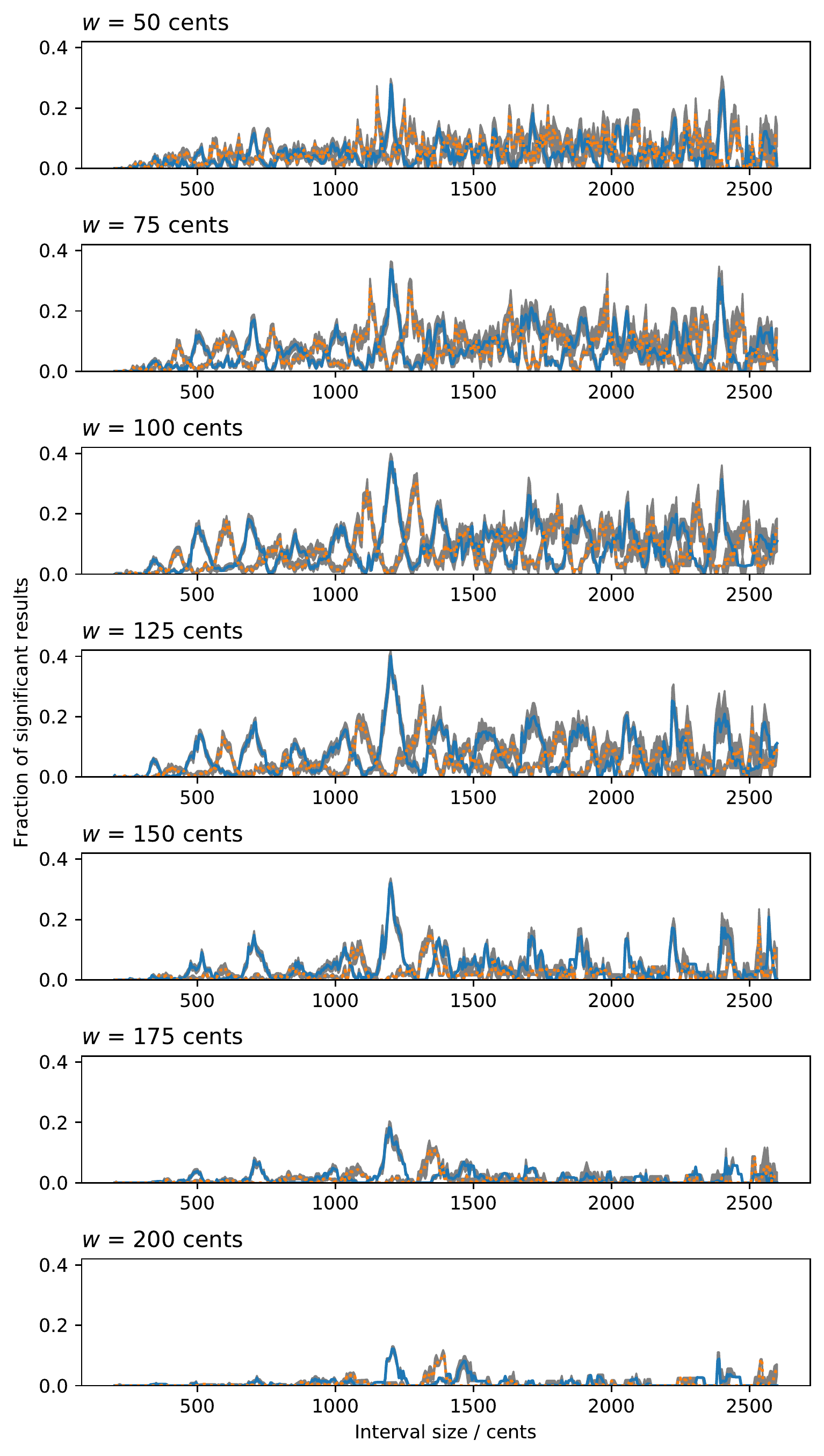}
  \caption{\label{fig:si6}
  Effect of changing $w$ on the fraction of significant results found
  for each interval size (see main text for details of statistical test).
  Blue line indicates that the interval is found significantly more than chance,
  while the orange line indicates the opposite.
  Shaded region shows bootstrapped $95\%$ confidence intervals.
  }
  \end{figure}

  \clearpage

  \begin{figure}[t!]  \centering
  \includegraphics[width=0.90\linewidth]{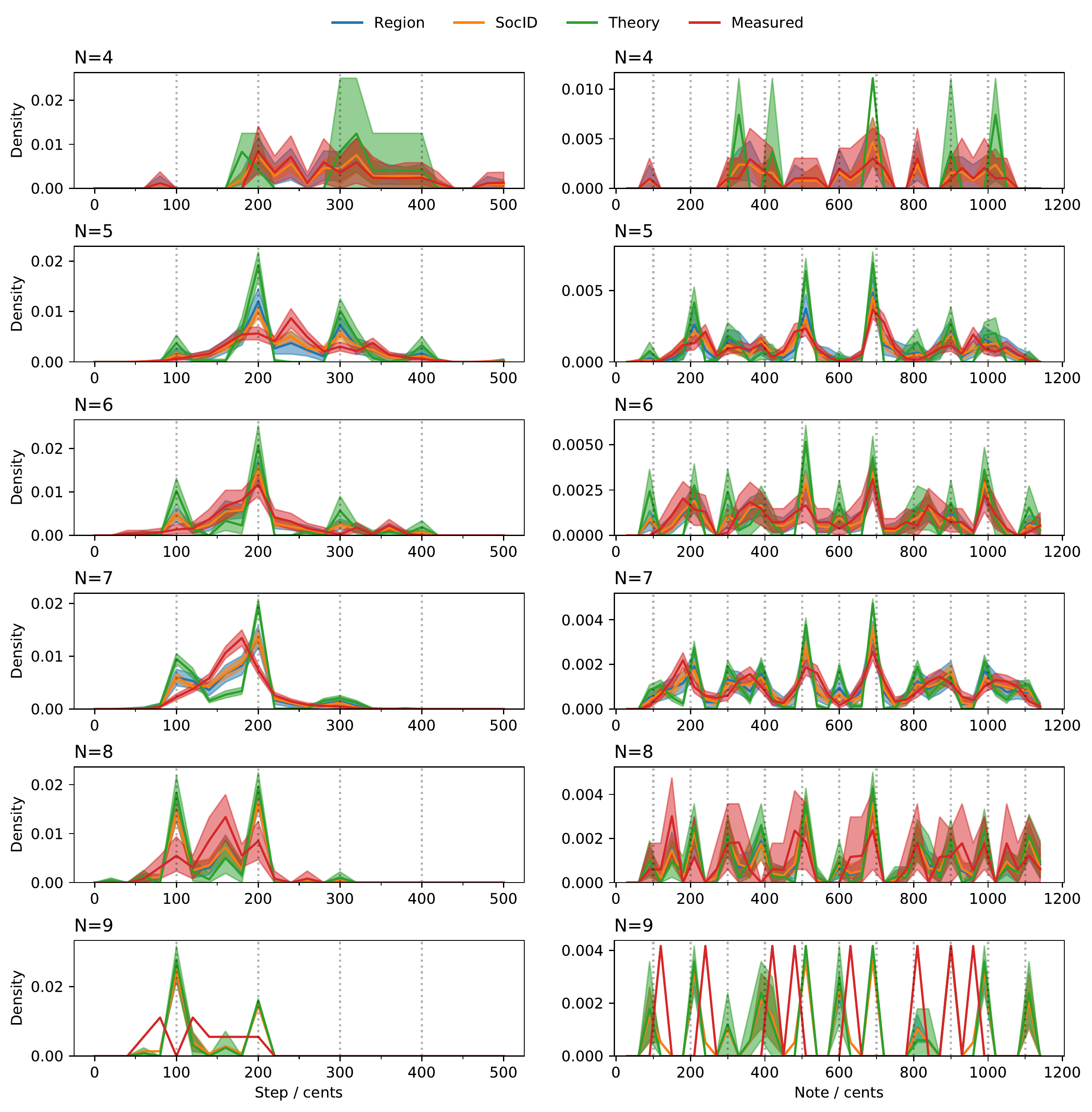}
  \caption{\label{fig:si7}
  Step size and scale note distributions for the SocID subsample (max $N_{samp}=5$),
  shown separately for sets of $N$-note scales. Separate lines are shown
  for alternative samples: Region (no more than 10 scales taken from each region);
  SocID (no more than 5 scales per society); only theory scales; only measured scales.
  Shaded region shows bootstrapped $95\%$ confidence intervals.
  }
  \end{figure}

  \clearpage

  \begin{figure}[t!]  \centering
  \includegraphics[width=0.95\linewidth]{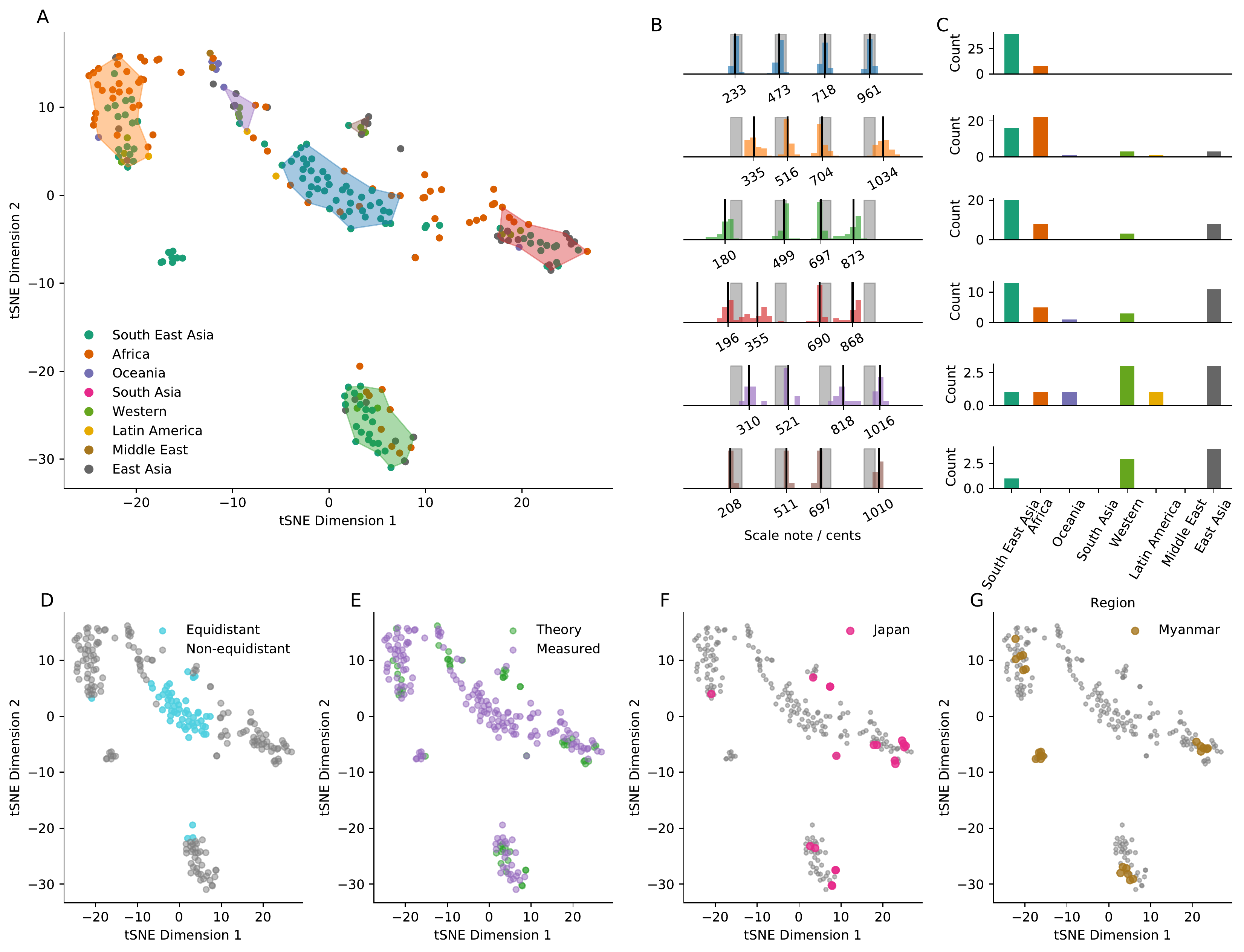}
  \caption{\label{fig:si8}
  Cross-cultural diversity of scales.
  A: 2-dimensional embedding of 232 pentatonic scales, with the six largest clusters
  indicated (shaded areas).
  B: Note distributions for each cluster. Black lines are shown for means of each note,
  and grey shading indicates equidistant scale notes $\pm$\SI{30}{cents}.
  C: Geographic distribution of each cluster.
  D-G: Embeddings are labelled with: equidistant scales (D; where notes are on average
  within \SI{30}{cents} of the equidistant values), theory vs measured scales (E),
  Japanese scales (F), Burmese scales (G).
  }
  \end{figure}

  \clearpage

  \begin{figure}[t!]  \centering
  \includegraphics[width=0.95\linewidth]{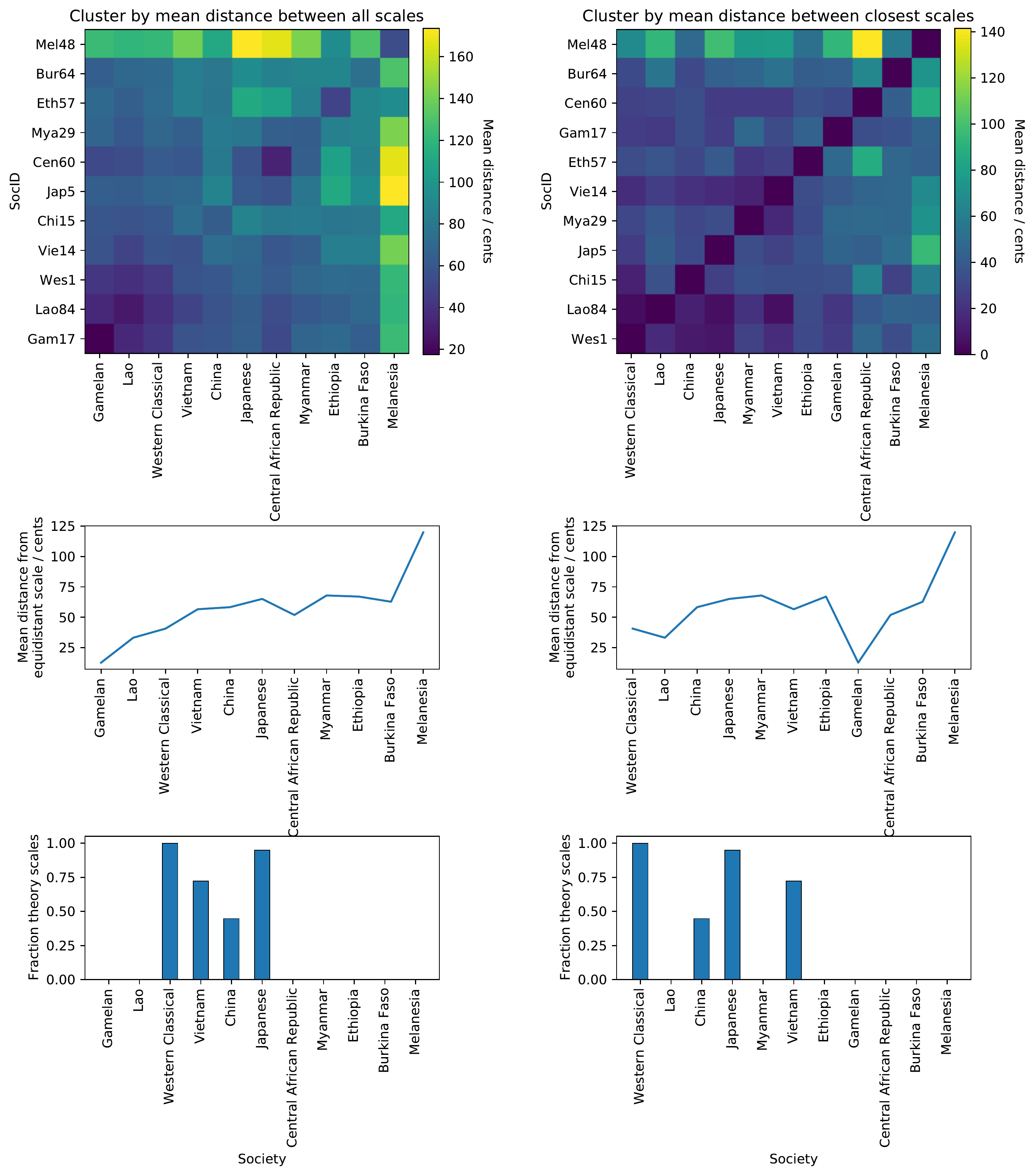}
  \caption{\label{fig:si11}
  A: Societies clustered by distance between 5-note scales for two distance metrics:
  (Left) mean distance between all pairs of scales between two clusters (symmetric distance metric: 
  distance from society A to society B is equal to the distance from society B to society A);
  clustering by the mean distance between all scales in society A, and the closest corresponding scale
  in society B (asymmetric distance metric).
  Societies are ordered according by hierarchical clustering according to each metric respectively, so
  that adjacent societies on the figure are more similar to each other.
  Societies are included in this analysis if there are at least 5 examples of scales.
  B: Mean distance from the equipentatonic scale.
  C: Fraction of scales found in the database that are theory scales.
  }
  \end{figure}

  \clearpage

  \begin{figure}[t!]  \centering
  \includegraphics[width=0.95\linewidth]{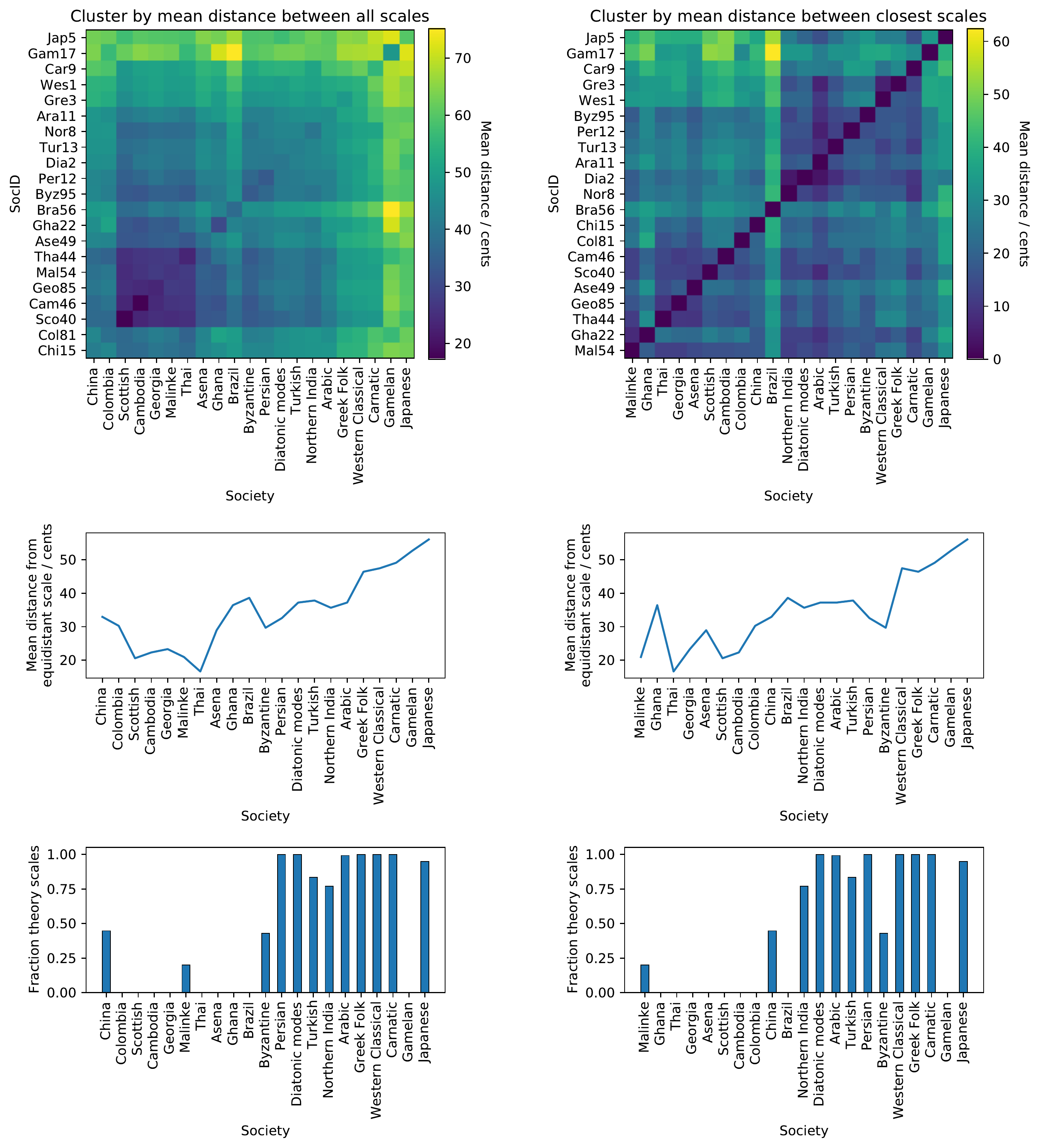}
  \caption{\label{fig:si12}
  A: Societies clustered by distance between 7-note scales for two distance metrics:
  (Left) mean distance between all pairs of scales between two clusters (symmetric distance metric: 
  distance from society A to society B is equal to the distance from society B to society A);
  clustering by the mean distance between all scales in society A, and the closest corresponding scale
  in society B (asymmetric distance metric).
  Societies are ordered according by hierarchical clustering according to each metric respectively, so
  that adjacent societies on the figure are more similar to each other.
  Societies are included in this analysis if there are at least 5 examples of scales.
  B: Mean distance from the equiheptatonic scale.
  C: Fraction of scales found in the database that are theory scales.
  }
  \end{figure}

  \clearpage

  \begin{figure}[t!]  \centering
  \includegraphics[width=0.95\linewidth]{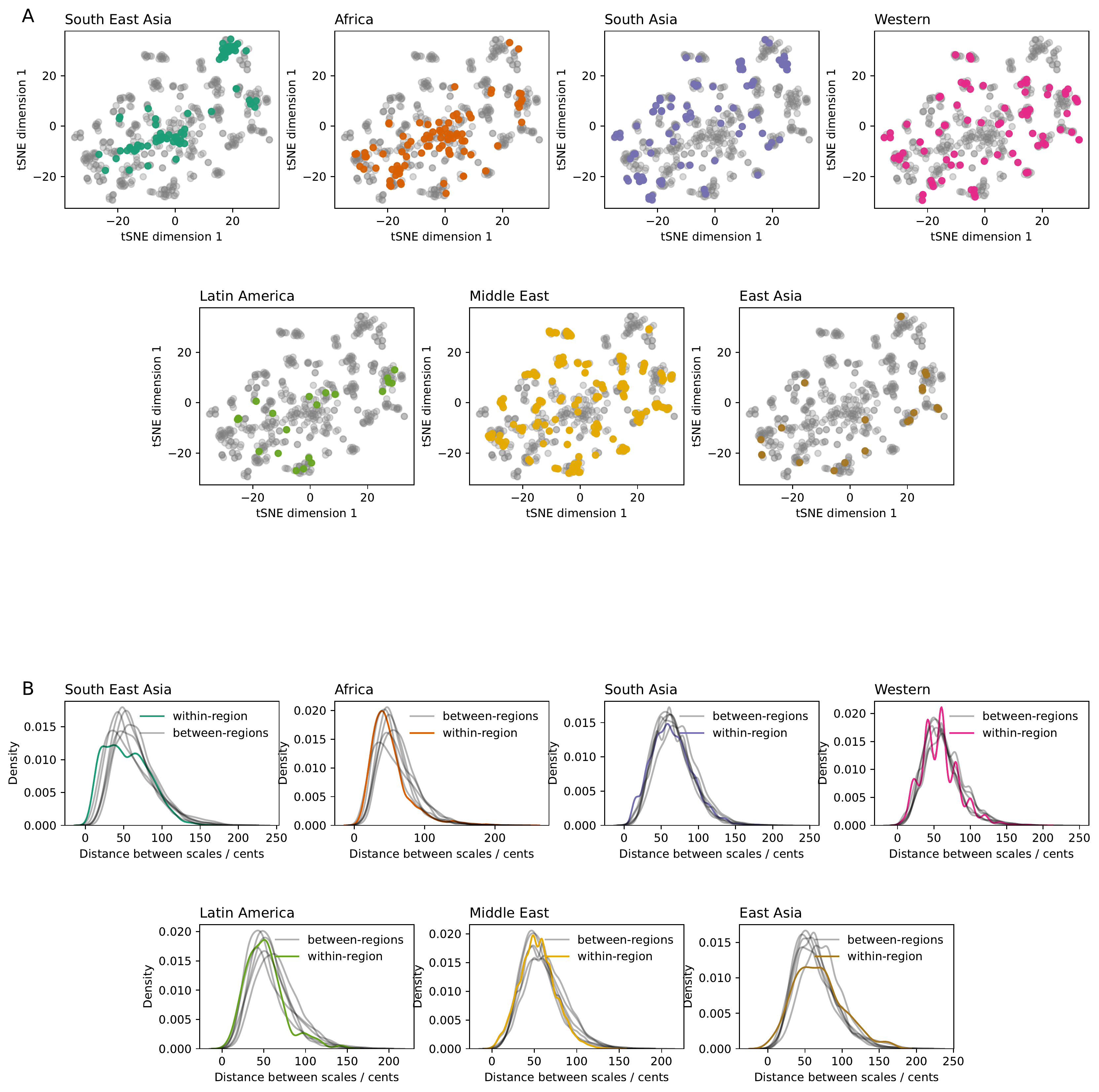}
  \caption{\label{fig:si4}
  A: tSNE embedding of 7-note scales, with different regions highlighted in each plot.
  Oceania was not included since there are too few examples in this case.
  B: Distributions of distances between all possible pairs of 7-note scales between
  one region and another (between-region distance) is shown in grey;
  distributions of distances between all pairs of scales within a region (within-region distance)
  is highlighted in each plot.
  }
  \end{figure}

  \clearpage

  \begin{figure}[t!]  \centering
  \includegraphics[width=0.90\linewidth]{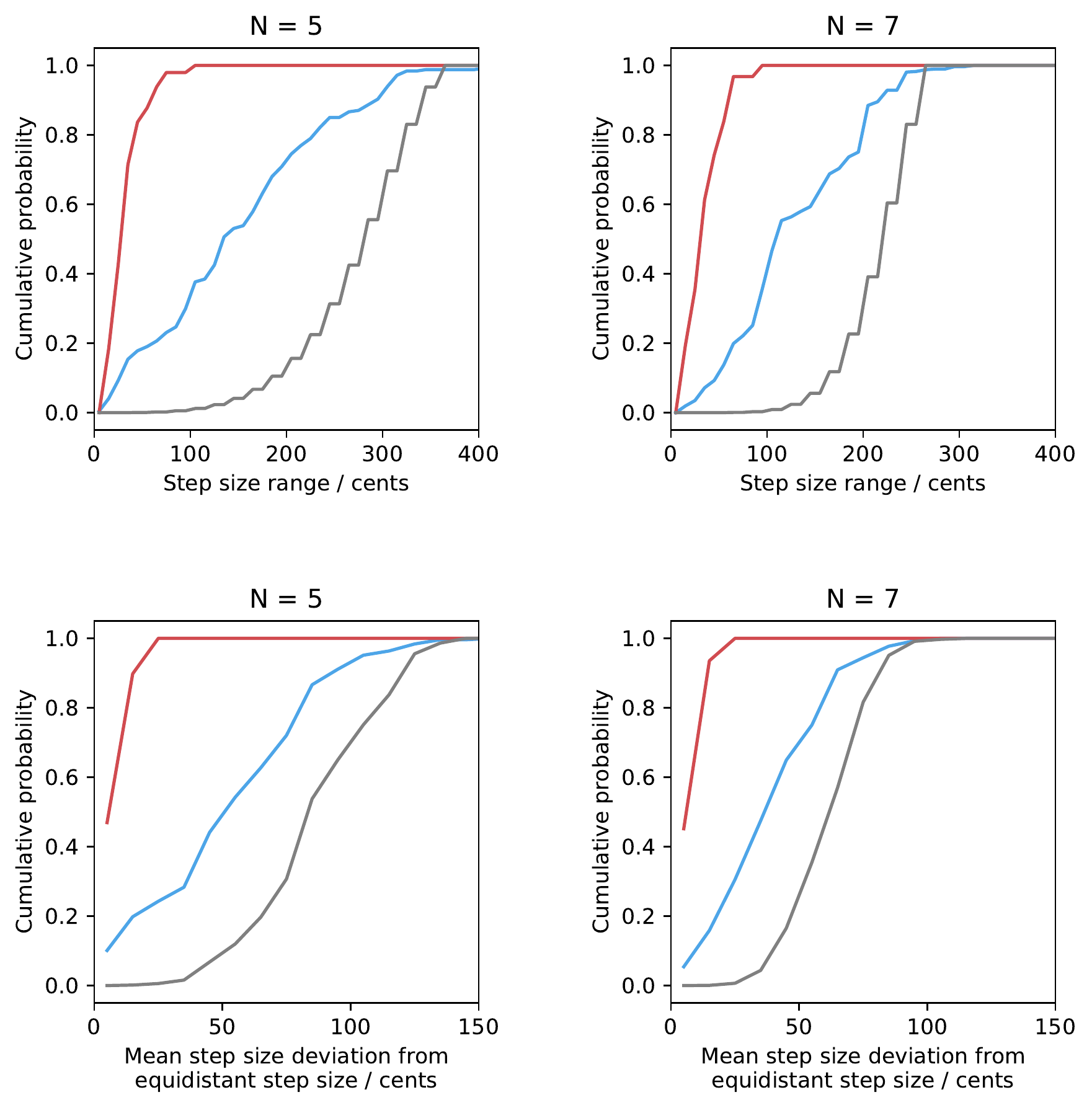}
  \caption{\label{fig:si9}
  Distributions of alternative metrics of equidistance for
  real scales, grid scales, Gamelan slendro scales ($N=5$), and Thai scales ($N=7$).
  Top: difference between smallest and largest step size.
  Bottom: Mean deviation of step sizes from equidistant step size.
  }
  \end{figure}

  \clearpage

  \begin{figure}[t!]  \centering
  \includegraphics[width=0.95\linewidth]{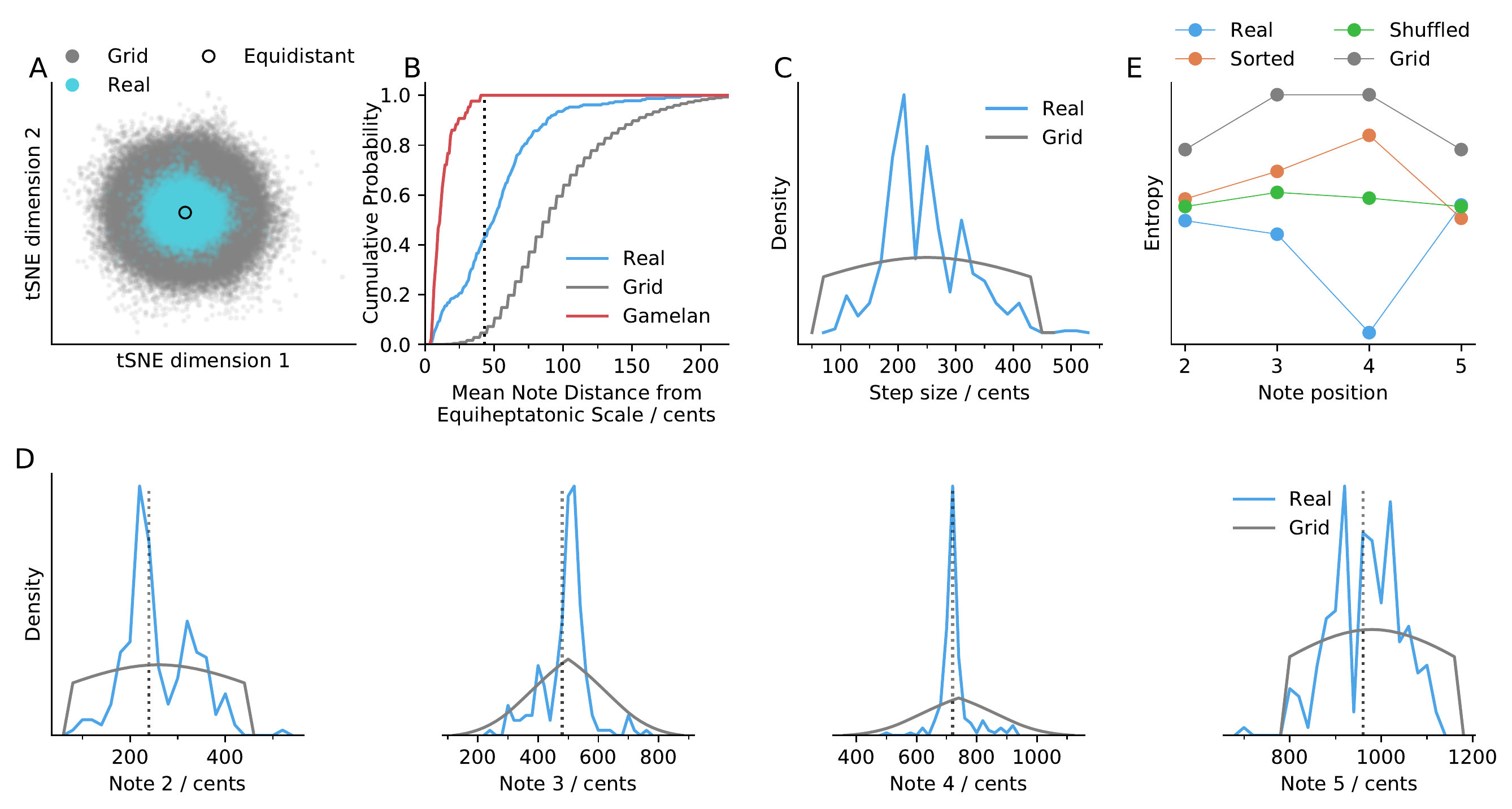}
  \caption{\label{fig:si10}
  Comparison of real scales with all possible 5-note scales enumerated on a grid (\SI{20}{cents} resolution; with step sizes limited to \SIrange{60}{320}{cents}).
  A: Two-dimensional embedding of grid scales. Grid scales that correspond to real scales (notes are on average within \SI{10}{cents} of a real scale) are highlighted cyan; a black circle shows the equidistant scale.
  B: Mean note distance from 5-note scales and the equipentatonic scale, for Thai scales, real scales, and all grid scales. 
  C: Step size histograms (bin size = \SI{20}{cents}) in real scales and grid scales.
  D: Scale note histograms (bin size = \SI{20}{cents}) for notes \numrange{2}{5} (no tonic and octave) for: real scales; real scales, but rearranged with their steps arranged in order of size (Sorted); real scales, but rearranged with their steps in random order (Shuffled); grid scales.
  E: Entropy of note distributions in D.
  }
  \end{figure}

  \clearpage

  \begin{figure}[t!]  \centering
  \includegraphics[width=0.90\linewidth]{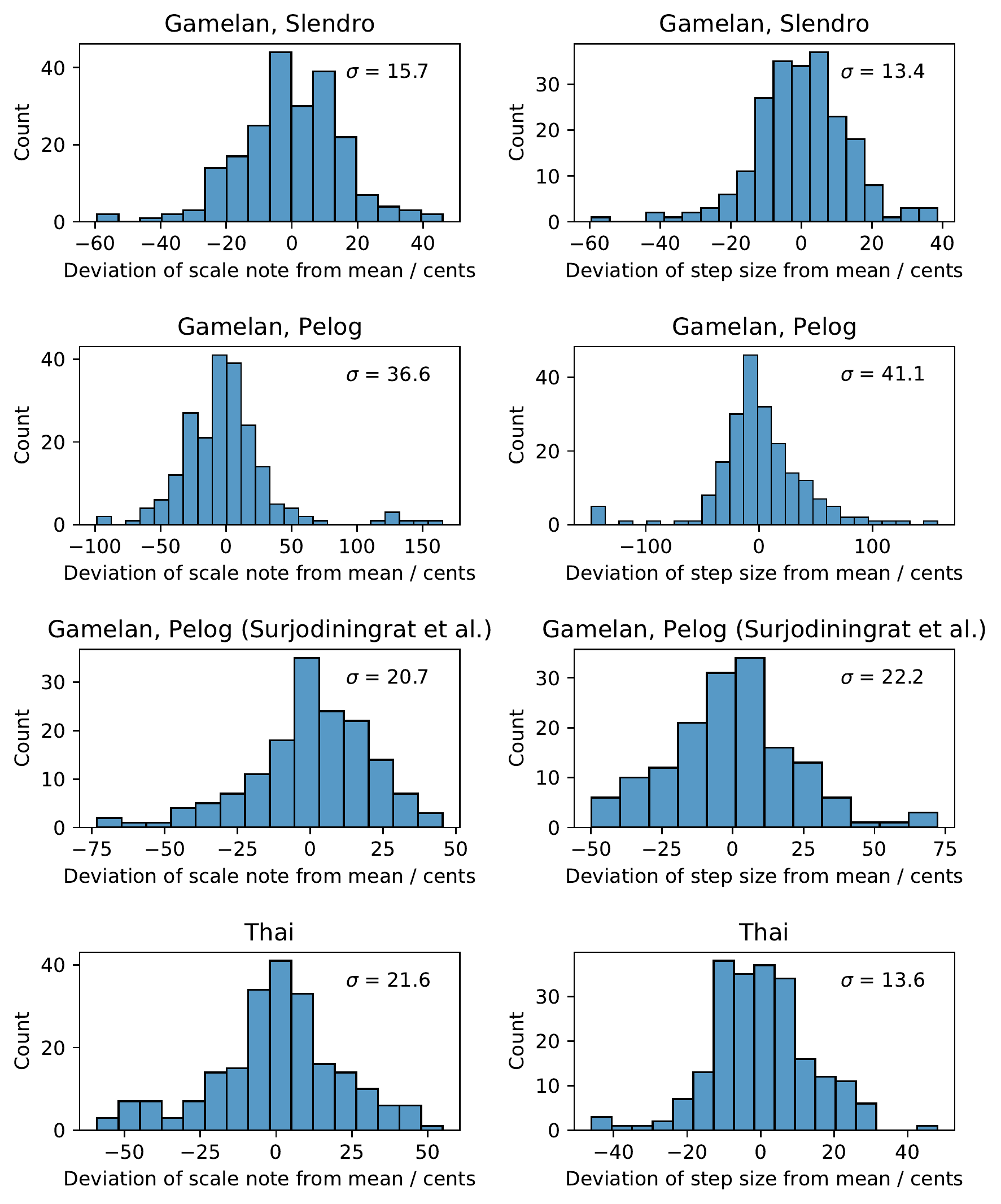}
  \caption{\label{fig:si13}
  Deviation of scale notes and step sizes from their mean values, for scales
  in the database: Gamelan \textit{slendro} scales (A), Gamelan \textit{pelog} scales (B),
  Gamelan \textit{pelog} scales from \citet{sur72} (C; this source accounts for 22 out of
  30 \textit{pelog} scales in the database), and Thai scales (D).
  Standard deviation is shown in text on each subplot. \\
  We are not exactly sure why the variance of \textit{pelog} scales is lower
  in Surjodiningra et al. than other sources, but it may have something to do with
  their attempt at only measuring tunings from `outstanding' gamelan orchestras.
  }
  \end{figure}

  \clearpage

  \begin{figure}[t!]  \centering
  \includegraphics[width=0.90\linewidth]{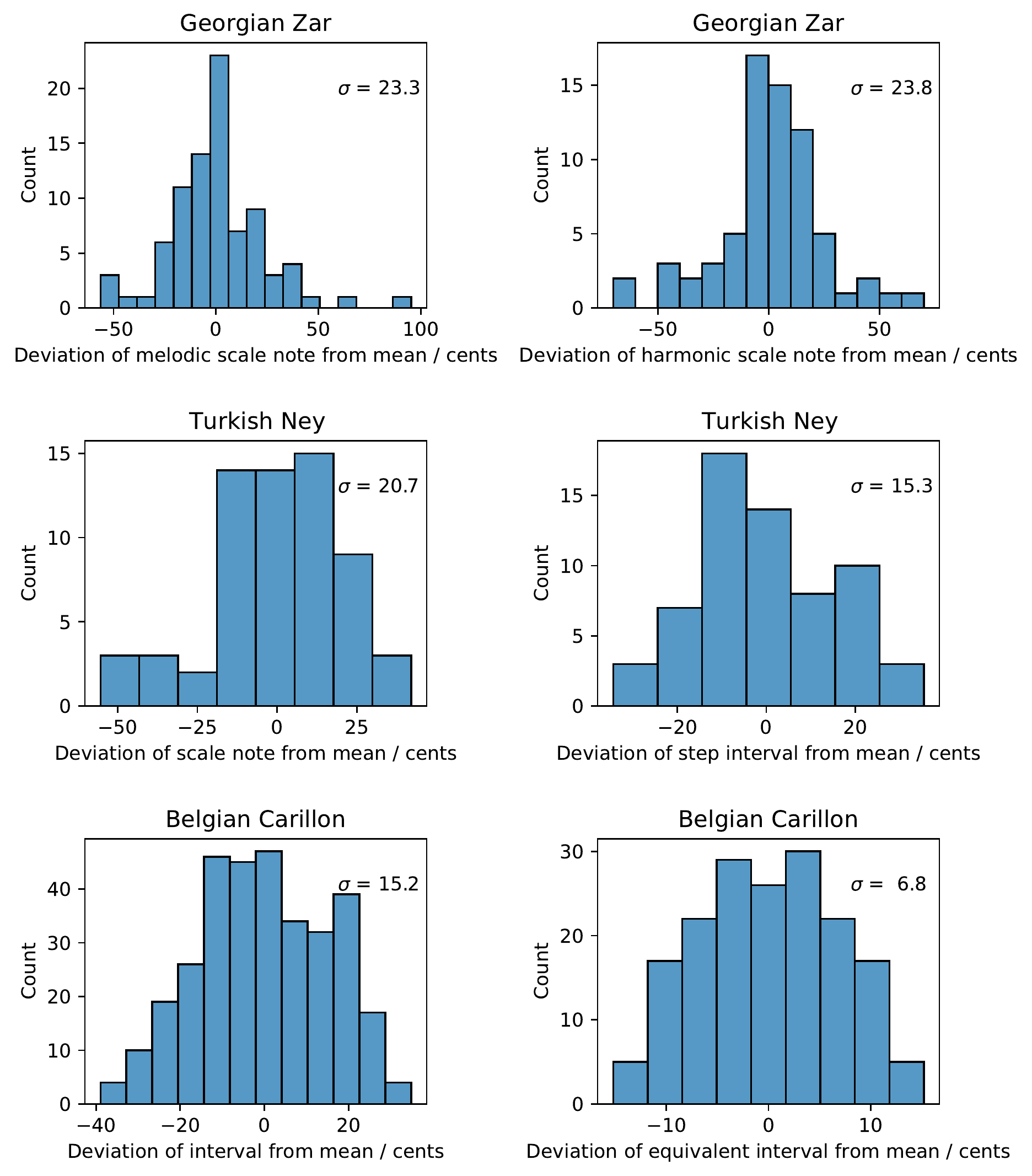}
  \caption{\label{fig:si14}
  Analysis of interval consistency for tuning information from three sources:
  Georgian polyphonic singing (A), from recordings of Zar (a type of funeral dirge)
  taken in different villages;~\cite{mzhmu20}
  a set of Turkish `ney' flutes (B);~\cite{tan11} a single Belgian carillon.~\cite{sch17}\\
  A: We align (in some villages certain notes are omitted) and compare all scale notes
  that were measured either from melodic pitch class histograms (left) or harmonic
  pitch class histograms (right). \\
  B: Since the flute tunings are already aligned (same number of notes; similar scale notes), we can calculate the
  deviation from mean for each note position, for scale notes (left) and step sizes (right). \\
  C: Unfortunately there is only one set of notes that spans \SI{2700}{cents}, so we performed
  two analyses of intervallic consistency within the instrument. We grouped all possible intervals
  by their distance to the nearest equal tempered (12-tet) interval, and calculated the deviation from
  the mean (left). It is not known whether the intended tuning is 12-tet (the authors speculate that
  mean-tempered tuning is used), so we also used a restricted set of intervals, where only intervals between
  the same pairs of notes were grouped together (right; \eg, C3-D3 is only grouped with C4-D4). This results in most
  groups only having two sets of measurements, which artefactually results in a lower standard deviation.
  It can be shown that using a sample size of 2 to estimate the standard deviation will typically
  result in a value that is approximately half of the actual standard deviation; thus, we believe that
  the first measurement (C, left) is a good approximation of the interval consistency of this instrument.
  }
  \end{figure}

  \clearpage

  \begin{table*}\centering
  \small
  \caption{\label{tab:data_source} The number of scales of each type from each source, in order:
  raw theory; raw measured from instrument; raw measured from recording; octave theory;
  octave instrument; octave recording; total octave scales.
  }
  \begin{tabular}{l|ccccccccc}
Ref & RefID & Year & T & M\textsubscript{inst} &  M\textsubscript{rec} & OT & OM\textsubscript{inst} &  OM\textsubscript{rec} & O\textsubscript{total} \\
  \toprule
\cite{rec18} & 2 & 2018 & 239 & 54 & 3 & 192 & 49 & 3 & 244 \\
\cite{hew13} & 1 & 2013 & 173 & 0 & 0 & 202 & 0 & 0 & 202 \\
\cite{sur72} & 12 & 1972 & 0 & 51 & 0 & 0 & 50 & 0 & 50 \\
\cite{vanam80} & 16 & 1980 & 0 & 31 & 0 & 0 & 48 & 0 & 48 \\
\cite{ell85} & 3 & 1885 & 25 & 29 & 0 & 21 & 22 & 0 & 43 \\
\cite{weiae13} & 40 & 2013 & 0 & 0 & 28 & 0 & 0 & 28 & 28 \\
\cite{garam75} & 44 & 1975 & 0 & 3 & 0 & 0 & 24 & 0 & 24 \\
\cite{min90} & 51 & 1990 & 0 & 9 & 0 & 0 & 17 & 0 & 17 \\
\cite{panjn13} & 60 & 2013 & 8 & 0 & 8 & 6 & 0 & 8 & 14 \\
\cite{garan15} & 34 & 2015 & 0 & 7 & 0 & 0 & 14 & 0 & 14 \\
\cite{hoam82} & 19 & 1982 & 14 & 0 & 0 & 13 & 0 & 0 & 13 \\
\cite{mil85} & 54 & 1985 & 0 & 3 & 0 & 0 & 13 & 0 & 13 \\
\cite{set05} & 30 & 2005 & 0 & 4 & 0 & 0 & 11 & 0 & 11 \\
\cite{mzhmu20} & 56 & 2020 & 0 & 0 & 22 & 0 & 0 & 11 & 11 \\
\cite{str09} & 32 & 2009 & 0 & 4 & 0 & 0 & 9 & 0 & 9 \\
\cite{mor76} & 13 & 1976 & 0 & 4 & 0 & 0 & 8 & 0 & 8 \\
\cite{bra67} & 6 & 1967 & 0 & 16 & 0 & 0 & 8 & 0 & 8 \\
\cite{zeme81} & 17 & 1981 & 0 & 8 & 0 & 0 & 7 & 0 & 7 \\
\cite{tan11} & 58 & 2011 & 0 & 0 & 9 & 0 & 0 & 7 & 7 \\
\cite{wis15} & 35 & 2015 & 0 & 4 & 0 & 0 & 6 & 0 & 6 \\
\cite{haee79} & 14 & 1979 & 0 & 9 & 0 & 0 & 6 & 0 & 6 \\
\cite{rousf69} & 8 & 1969 & 0 & 3 & 0 & 0 & 6 & 0 & 6 \\
\cite{anija82} & 18 & 1982 & 0 & 2 & 0 & 0 & 6 & 0 & 6 \\
\cite{yuapp87} & 23 & 1987 & 0 & 4 & 0 & 0 & 6 & 0 & 6 \\
\cite{kubam64} & 5 & 1964 & 0 & 5 & 0 & 0 & 6 & 0 & 6 \\
\cite{keemp91} & 24 & 1991 & 0 & 0 & 5 & 0 & 0 & 5 & 5 \\
\cite{kus10} & 33 & 2010 & 0 & 4 & 1 & 0 & 4 & 1 & 5 \\
\cite{kubam80} & 15 & 1980 & 0 & 5 & 0 & 0 & 5 & 0 & 5 \\
\cite{kun67} & 7 & 1967 & 0 & 3 & 8 & 0 & 0 & 5 & 5 \\
\cite{jonam60} & 46 & 1960 & 0 & 10 & 0 & 0 & 5 & 0 & 5 \\
\cite{thram78} & 50 & 1978 & 0 & 5 & 0 & 0 & 5 & 0 & 5 \\
\cite{zhaa04} & 29 & 2004 & 0 & 5 & 0 & 0 & 4 & 0 & 4 \\
\cite{kim76} & 38 & 1976 & 0 & 0 & 4 & 0 & 0 & 4 & 4 \\
\cite{bad19} & 37 & 2019 & 0 & 3 & 0 & 0 & 4 & 0 & 4 \\
\cite{kni76} & 11 & 1976 & 0 & 4 & 0 & 0 & 4 & 0 & 4 \\
\cite{mcnja08} & 31 & 2008 & 0 & 1 & 0 & 0 & 3 & 0 & 3 \\
\cite{got85} & 22 & 1986 & 0 & 4 & 0 & 0 & 3 & 0 & 3 \\
\cite{nammp98} & 39 & 1998 & 0 & 0 & 3 & 0 & 0 & 3 & 3 \\
\cite{kubwm92} & 49 & 1992 & 0 & 1 & 0 & 0 & 3 & 0 & 3 \\
\cite{men12} & 59 & 2012 & 0 & 3 & 0 & 0 & 3 & 0 & 3 \\
\cite{sche01} & 27 & 2001 & 0 & 2 & 0 & 0 & 2 & 0 & 2 \\
\cite{carja93} & 26 & 1993 & 0 & 2 & 0 & 0 & 2 & 0 & 2 \\
\cite{traam71} & 10 & 1971 & 0 & 2 & 0 & 0 & 2 & 0 & 2 \\
\cite{traam70} & 9 & 1970 & 0 & 2 & 0 & 0 & 2 & 0 & 2 \\
\cite{traam91} & 25 & 1991 & 0 & 1 & 0 & 0 & 2 & 0 & 2 \\
\cite{wacna50} & 4 & 1950 & 0 & 2 & 0 & 0 & 2 & 0 & 2 \\
\cite{kubam84} & 20 & 1984 & 0 & 1 & 0 & 0 & 2 & 0 & 2 \\
\cite{kubyt85} & 21 & 1985 & 0 & 2 & 0 & 0 & 2 & 0 & 2 \\
\cite{kubam63} & 48 & 1963 & 0 & 1 & 0 & 0 & 2 & 0 & 2 \\
\cite{sunst69} & 52 & 1969 & 0 & 2 & 0 & 0 & 2 & 0 & 2 \\
\cite{copef18} & 53 & 2018 & 0 & 0 & 2 & 0 & 0 & 2 & 2 \\
\cite{moric18} & 36 & 2018 & 0 & 1 & 0 & 0 & 1 & 0 & 1 \\
\cite{attpi04} & 28 & 2004 & 0 & 2 & 0 & 0 & 1 & 0 & 1 \\
\cite{jonet66} & 45 & 1966 & 0 & 1 & 0 & 0 & 1 & 0 & 1 \\
\cite{kubam62} & 47 & 1962 & 0 & 2 & 0 & 0 & 1 & 0 & 1 \\
\cite{sch20} & 55 & 2020 & 0 & 0 & 9 & 0 & 0 & 1 & 1 \\
\cite{mokna37} & 57 & 1937 & 0 & 1 & 0 & 0 & 1 & 0 & 1 \\
\cite{ambwm05} & 41 & 2005 & 0 & 0 & 2 & 0 & 0 & 0 & 0 \\
\cite{ambms06} & 42 & 2006 & 0 & 0 & 7 & 0 & 0 & 0 & 0 \\
\cite{cooyi71} & 43 & 1971 & 0 & 1 & 0 & 0 & 0 & 0 & 0 \\
  \toprule
 &  & \textbf{Total} & 459 & 323 & 111 & 434 & 384 & 78 & 896 \\
\end{tabular}
\end{table*}

  \clearpage

  \begin{table*}\centering
  \caption{\label{tab:tunings}
    The tunings used for `theory' scales depending on culture.
  }
  \begin{tabular}{lcccccccc}
    Society &  12-tet & 24-tet & 53-tet & Just Intonation & Pythagorean & Persian & Turkish & Shi-er-lu  \\
  \midrule
    Western Classical         & X &   &   &   &   &   &   &   \\
    Jazz                      & X &   &   &   &   &   &   &   \\
    Diatonic modes            & X &   &   & X & X &   &   &   \\
    Greek Folk                & X &   &   &   &   &   &   &   \\
    Japanese                  &   &   &   &   & X &   &   &   \\
    Malink\'{e}               &   &   &   &   & X &   &   &   \\
    Chinese                   &   &   &   &   &   &   &   & X \\
    Hindustani                & X &   &   & X &   &   &   &   \\
    Carnatic                  &   &   &   & X &   &   &   &   \\
    Arabian                   &   & X & X &   &   &   &   &   \\
    Persian                   &   &   &   &   &   & X &   &   \\
    Turkish                   &   &   & X &   &   &   & X &   \\
  \bottomrule
  \end{tabular}
  \end{table*}

  \clearpage

\bibliographystyle{unsrtnat}
\bibliography{world_scale}